\documentclass[preprint,prd,aps,showpacs,showkeys,nofootinbib]{revtex4-1}
\usepackage{amssymb}
\usepackage{amsmath}

\usepackage{graphicx}
\usepackage{subfigure}

\usepackage{float}

\begin{document}

\title{Gauss-Bonnet inflation with a constant rate of roll}

\author{Tie-Jun Gao$^1$}
\email{tjgao@xidian.edu.cn}

\affiliation{$^1$School of Physics and Optoelectronic Engineering, Xidian University, Xi'an 710071, China}

\begin{abstract}
We consider the constant-roll condition in the model of the inflaton nonminimal coupling to the Gauss-Bonnet term. By assuming the first Gauss-Bonnet flow parameter $\delta_1$ is a constant, we discuss the constant-roll inflation with constant $\epsilon_1$, constant $\epsilon_2$ and constant $\eta_H$, respectively. Using the Bessel function approximation, we get the analytical expressions for the scalar and tensor power spectrum  and  derive the scalar spectral index  $n_{\mathcal{R}}$ and the tensor to scalar ratio $r$ to the first order of $\epsilon_1$.  By using the Planck 2018 observations constraint on $n_{\mathcal{R}}$ and $r$,  we obtain some feasible parameter space and show the result on the $n_{\mathcal{R}}-r$ region. The scalar potential  is also reconstructed in some spectral cases.
\end{abstract}

\keywords{Constant-roll,  Gauss-Bonnet term,  Inflation} \pacs{}

\maketitle

\section{Introduction\label{sec1}}
\indent
Inflation in the early Universe has become a well established part of modern cosmology, it naturally generates
the density perturbations that become the seeds of large-scale structure and temperature anisotropies of the cosmic microwave background(CMB). The predictions have been confirmed by numerous observations, such as the WMAP \cite{ref1} and Planck space missions \cite{ref2}.  Recently, the Planck 2018 data constrain the scalar spectral index and the tensor-to-scalar ratio to be $n_{\mathcal{R}}=0.9649\pm0.0042$ at $68\%$ confidence level and $r_{0.002}<0.10$ at $95\%$ confidence level \cite{ref2},  which will narrow down some inflationary models.

In the conventional inflationary models, it is usually assumed that the scalar field rolls down slowly, or more precisely to assumed that the flow parameters satisfy the conditions $\epsilon_H\ll1$ and $\eta_H\ll1$. This  so called slow-roll inflation leads to a nearly scale-invariant spectrum of density perturbations and are consistent with the latest observations.
Recently, a new type of inflation called constant-roll inflation has been put forward in \cite{ref3}, where one of the flow parameters $\eta_H$ is assumed to be a constant, and not necessarily small. Such class of models can also produce a scale-invariant spectrum and is compatible with the observational constraints without the assumption of slow-roll approximation. The constant-roll condition with small  $\eta_H$ is compatible with the slow-roll inflation, and a special value $\eta_H=3$ amounts to the ultra-slow-roll inflation \cite{ref4,ref5}, in which the potential is very flat and the inflaton almost stops rolling.The constant-roll inflationary models are discussed in more details in \cite{ref6,ref7,ref8}, and extended to modified theories of gravity\cite{ref9,ref10,ref11,ref12,ref13,ref14} or two fields case\cite{ref15}, and using other constant-roll conditions \cite{ref16,ref17,ref18}.  Other developments are  appeared in \cite{ref19,ref20,ref21}.

Since the inflation occurs in the early Universe which is believed to be described by quantum gravity, so it is interest to discuss inflation in the framework of quantum gravity theories, such as string theory. It is known that when discuss the  effective action in the early universe, the  correction terms of higher orders in the curvature coming from superstrings may play a significant role,
and the simplest of such correction is the Gauss-Bonnet (GB) term in the low-energy effective action of the heterotic string\cite{ref22,ref23}. There are many works discussing accelerating cosmology with the GB correction in four and higher dimensions\cite{ref24,ref25,ref26,ref27}, and with nonminimal coupled to the inflaton\cite{ref28,ref29,ref30} or to the Higgs\cite{ref31,ref32,ref33,ref34,ref35}.

In this work, we shall discuss the constant-roll inflation in the model with the inflation field $\phi$ nonminimal couple to the GB term. The presence of such coupling will generate a new degree of freedom, so here we consider a special condition that the first GB flow parameter $\delta_1$ is a constant.
Assuming that one of the flow parameters, $\epsilon_1$, $\epsilon_2$ or  $\eta_H$ is constant,  we derive the analytical expressions of the scalar spectral index $n_{\mathcal{R}}$ and the tensor to scalar ratio  $r$ for each case. Combine with the Planck 2018 observations, we obtain some feasible parameter spaces, and find that in some cases, the chosen of GB flow parameter $\delta_1>1$ can also produce a nearly scale-invariant scalar power spectrum, which is different from the slow-roll inflation. At last, we also reconstruct the potential in some spectral cases.

The outline of this paper is as follows: In the next section, we briefly review the inflationary model with nonminimal coupling to the Gauss-Bonnet term. In Section 3,4 and 5, we focus on the inflationary models with constant $\epsilon_1$, constant $\epsilon_2$ or with constant $\eta_H$, respectively, derive the formalism for the scalar spectral index and the tensor-to-scalar ratio of each model and constrain the parameter space with the Planck 2018 data. In Section 6, we try to reconstruct the potential in some spectral cases. And the final section is devoted to summary.









\section{Inflation with a Gauss-Bonnet coupling \label{sec2}}
In this section, we shall review the inflation with the GB coupling, and present the mode equations of scalar and tensor perturbations.
 Consider the following action with the inflation field $\phi$ coupled to the GB term
\begin{eqnarray}
&&S=\int d^{4} x \sqrt{-g}\left[\frac{1}{2} R-\frac{\omega}{2}(\nabla \phi)^{2}-V(\phi)-\frac{1}{2} \xi(\phi) R_{\mathrm{GB}}^{2}\right],
\label{infp9}
\end{eqnarray}
where $V(\phi)$ is the potential for the scalar field $\phi$,  $R_{\mathrm{GB}}^{2}=R_{\mu\nu\rho\sigma}R^{\mu\nu\rho\sigma}-4R_{\mu\nu}R^{\mu\nu}+R^2$ is the GB term, and $\xi(\phi)$ is the coupling  functional. The coefficient $\omega$ in the kinetic term can be chosen the values $\pm1$ to ensure conventional inflationary behaviour\cite{ref28,ref29}. We work in Planckian units, $\hbar=c=8 \pi G=1$.

In the Friedmann-Robertson-Walker homogeneous universe, the background equations can be written as
\begin{eqnarray}
&&6 H^{2}=\omega \dot{\phi}^{2}+2 V+24 \dot{\xi} H^{3}, \\
&&2 \dot{H}=-\omega \dot{\phi}^{2}+4 \ddot{\xi} H^{2}+4 \dot{\xi} H\left(2 \dot{H}-H^{2}\right), \\
&&\omega(\ddot{\phi}+3 H \dot{\phi})+V_{, \phi}+12 \xi_{, \phi} H^{2}\left(\dot{H}+H^{2}\right)=0,
\label{infp9}
\end{eqnarray}
where a dot represents a derivative with respect to cosmic time and $(...)_{,\phi}$ denotes a derivative
with respect to the field $\phi$.

In standard inflation, it is useful to define a series of flow parameters, such as the Hubble flow parameters
\begin{eqnarray}
&&\epsilon_{H}=-\frac{\dot{H}}{H^{2}}, \quad \eta_{H}=-\frac{\ddot{H}}{2H\dot{H}},
\label{infp9}
\end{eqnarray}
furthermore, we also introduce the horizon flow parameters
\begin{eqnarray}
&&\epsilon_{1}=-\frac{\dot{H}}{H^{2}}, \quad \epsilon_{i+1}=\frac{\dot{\epsilon_{i}}}{H \epsilon_{i}}, \quad i \geq 1.
\label{infp9}
\end{eqnarray}
In the presence of the GB coupling, the new degrees of freedom suggest to defining another hierarchy of flow parameters\cite{ref29}
\begin{eqnarray}
&&\delta_{1}=4 \dot{\xi} H, \quad \delta_{i+1}=\frac{\dot{\delta_{i}}}{H \delta_{i}}, \quad i \geq 1.
\label{infp9}
\end{eqnarray}
For the slow-roll inflation, these flow parameters should satisfy the slow-roll conditions $|\epsilon_i|\ll1$ and $|\delta_i|\ll1$.
However, in the constant-roll inflation, such conditions are not necessary to satisfy, we only need $0<\epsilon_1<1$ to achieve a significative inflation.

At linear order in perturbation theory, the Fourier modes of curvature perturbations satisfy the Mukhanov-Sasaki equation
\begin{eqnarray}
&&v^{\prime \prime}+\left(c_{\mathcal{R}}^{2} k^{2}-\frac{z_{\mathcal{R}}^{\prime \prime}}{z_{\mathcal{R}}}\right) v=0,
\label{infp9}
\end{eqnarray}
where a prime represents the  derivative with respect to conformal time $\tau$. $z_{\mathcal{R}}$ and the sound speed $c_{\mathcal{R}}$ can be expressed in terms of the horizon  and GB flow parameters as
\begin{equation}
z_{\mathcal{R}}^{2}=a^{2} \frac{F}{\left(1-\frac{1}{2} \Delta\right)^{2}},
\end{equation}
\begin{equation}
c_{\mathcal{R}}^{2}=1-\Delta^{2} \frac{2 \epsilon_{1}+\frac{1}{2} \delta_{1}\left(1-5 \epsilon_{1}-\delta_{2}\right)}{F},
\end{equation}
with $\Delta=\delta_{1} /\left(1-\delta_{1}\right)$ and $F \equiv 2 \epsilon_{1}-\delta_{1}\left(1+\epsilon_{1}-\delta_{2}\right)+\frac{3}{2} \Delta \delta_{1}$.
And the effective mass term reads
\begin{equation}\begin{aligned}
\frac{z_{\mathcal{R}}^{\prime \prime}}{z_{\mathcal{R}}}=& a^{2} H^{2}\left[2-\epsilon_{1}+\frac{3}{2} \frac{\dot{F}}{H F}+\frac{3}{2} \frac{\dot{\Delta}}{H\left(1-\frac{1}{2} \Delta\right)}+\frac{1}{2} \frac{\ddot{F}}{H^{2} F}+\frac{1}{2} \frac{\ddot{\Delta}}{H^{2}\left(1-\frac{1}{2} \Delta\right)}\right.\\
&\left.-\frac{1}{4} \frac{\dot{F}^{2}}{H^{2} F^{2}} +\frac{1}{2} \frac{\dot{\Delta}^{2}}{H^{2}\left(1-\frac{1}{2} \Delta\right)^{2}}+\frac{1}{2} \frac{\dot{\Delta}}{H\left(1-\frac{1}{2} \Delta\right)} \frac{\dot{F}}{H F}\right],
\end{aligned}\end{equation}
with
\begin{equation}\begin{aligned}
\frac{\dot{F}}{H}=& \epsilon_{1} \epsilon_{2}\left(2-\delta_{1}\right)-\delta_{1} \delta_{2}\left(1+\epsilon_{1}-\delta_{2}-\delta_{3}\right)+\frac{3}{2} \Delta \delta_{2}\left(\Delta+\delta_{1}\right), \\
\frac{\dot{\Delta}}{H}=& \Delta^{2} \frac{\delta_{2}}{\delta_{1}}, \\
\ddot{H} =&\epsilon_{1} \epsilon_{2}\left(-\epsilon_{1}+\epsilon_{2}+\epsilon_{3}\right)\left(2-\delta_{1}\right)+\epsilon_{1} \delta_{1} \delta_{2}\left(1+\epsilon_{1}-2 \epsilon_{2}-\delta_{2}-\delta_{3}\right)\\
&-\delta_{1} \delta_{2}^{2}\left(1+\epsilon_{1}-\delta_{2}-\delta_{3}\right) -\delta_{1} \delta_{2} \delta_{3}\left(1+\epsilon_{1}-2 \delta_{2}-\delta_{3}-\delta_{4}\right) \\
&+\frac{3}{2} \Delta \delta_{2}\left(\Delta+\delta_{1}\right)\left(-\epsilon_{1}+\Delta \frac{\delta_{2}}{\delta_{1}}+\delta_{3}\right) +\frac{3}{2} \Delta \delta_{2}\left(\Delta^{2} \frac{\delta_{2}}{\delta_{1}}+\delta_{1} \delta_{2}\right), \\
\ddot{\Delta} =&\Delta^{2} \frac{\delta_{2}}{\delta_{1}}\left(-\epsilon_{1}+2 \Delta \frac{\delta_{2}}{\delta_{1}}-\delta_{2}+\delta_{3}\right).
\end{aligned}\end{equation}
Similarly, for the tensor perturbations, the Fourier modes satisfy
\begin{eqnarray}
&&u^{\prime \prime}+\left(c_{T}^{2} k^{2}-\frac{z_{T}^{\prime \prime}}{z_{T}}\right) u=0,
\label{infp9}
\end{eqnarray}
with
\begin{equation}
z_{T}^{2}=a^{2}\left(1-\delta_{1}\right),
\end{equation}
\begin{equation}
c_{T}^{2}=1+\Delta\left(1-\epsilon_{1}-\delta_{2}\right).
\end{equation}
And the effective mass term in the tensor mode can be written in terms of the flow parameters
\begin{equation}\begin{aligned}
\frac{z_{T}^{\prime \prime}}{z_{T}}=& a^{2} H^{2}\left[2-\epsilon_{1}-\frac{3}{2} \Delta \delta_{2}-\frac{1}{2} \Delta \delta_{2}\left(-\epsilon_{1}+\delta_{2}+\delta_{3}\right)-\frac{1}{4} \Delta^{2} \delta_{2}^{2}\right].
\end{aligned}\end{equation}

In the following sections, we shall focus on the constant-roll inflation in the model with nonminimal coupling to the GB term.



\section{Constant-roll inflation with constant $\epsilon_1$  \label{sec3}}
We focus on the constant-roll inflation in the model with nonminimal coupling to the GB term. The presence of such coupling will generate a new degree of freedom $\xi(\phi)$, so in this work we consider a special condition that the first GB flow parameter $\delta_1$ is also a constant, then from the definition of GB flow parameters (7) we have $\delta_i=0(i\geq 2)$.

In the following, we first discuss the case with $\epsilon_1=$ constant. From the relations
\begin{equation}
\frac{d}{d \tau}\left(\frac{1}{a H}\right)=-1+\epsilon_{1},
\end{equation}
and assuming that  $\epsilon_1$ is a constant, one can express the factor $aH$ in (11) and (16) as a function of conformal time $\tau$
\begin{equation}
aH=-\frac{1}{\tau }\left(\frac{1}{1-\epsilon_1}\right),
\end{equation}
then
\begin{equation}
\nu_{\mathcal{R}}^2=\tau^2\frac{z_{\mathcal{R}}^{\prime \prime}}{z_{\mathcal{R}}}+\frac{1}{4},
\end{equation}
can be approximated as a constant. Therefore the general solution to the mode equation (8) is a linear
combination of Hankel functions of order $\nu_{\mathcal{R}}$
\begin{equation}
v=\frac{\sqrt{\pi|\tau|}}{2} e^{i\left(1+2 \nu_{\mathcal{R}}\right) \pi / 4}\left[c_{1} H_{\nu_{\mathcal{R}}}^{(1)}\left(c_{\mathcal{R}} k|\tau|\right)+c_{2} H_{\nu_{\mathcal{R}}}^{(2)}\left(c_{\mathcal{R}} k|\tau|\right)\right]
\end{equation}
Choosing $c_1=1$ and $c_2=0$, the usual Minkowski vacuum state is recovered in the asymptotic past. Then
on super-horizon scales, $c_{\mathcal{R}}k\ll aH$, the power spectrum of the scalar perturbation is
\begin{equation}\begin{aligned}
\mathcal{P}_{\mathcal{R}} &=\frac{c_{\mathcal{R}}^{-3}}{F} \frac{H^{2}}{4 \pi^{2}}\left(\frac{1-\Delta / 2}{a H|\tau|}\right)^{2} \frac{\Gamma^{2}\left(\nu_{\mathcal{R}}\right)}{\Gamma^{2}(3 / 2)}\left(\frac{c_{\mathcal{R}} k|\tau|}{2}\right)^{3-2 \nu_{\mathcal{R}}} \\
&\left.\simeq \frac{2^{2 \nu_{\mathcal{R}}-3} c_{\mathcal{R}}^{-3}}{F} \frac{H^{2}}{4 \pi^{2}} \frac{\Gamma^{2}\left(\nu_{\mathcal{R}}\right)}{\Gamma^{2}(3 / 2)}\left(1-\frac{\Delta}{2}\right)^2\left(\frac{1}{1-\epsilon_1}\right)^{1-2\nu_{\mathcal{R}}}\right|_{c_{\mathcal{R}} k=a H},
\end{aligned}\end{equation}
with the scalar spectral index is
\begin{equation}
n_{\mathcal{R}}-1=\frac{d\ln \mathcal{P}_{\mathcal{R}}}{d\ln k}=3-2\nu_{\mathcal{R}}.
\end{equation}

Using the same procedure as in the case of scalar perturbations, we get the power spectrum of tensor perturbations
\begin{equation}\begin{aligned}
\mathcal{P}_{T} &=\frac{8 c_{T}^{-3}}{1-\delta_{1}} \frac{H^{2}}{4 \pi^{2}}\left(\frac{1}{a H|\tau|}\right)^{2} \frac{\Gamma^{2}\left(\nu_{T}\right)}{\Gamma^{2}(3 / 2)}\left(\frac{c_{T} k|\tau|}{2}\right)^{3-2 \nu_{T}} \\
&\left.\simeq 2^{2 \nu_{T}} c_{T}^{-3} \frac{H^{2}}{4 \pi^{2}} \frac{\Gamma^{2}\left(\nu_{T}\right)}{\Gamma^{2}(3 / 2)}\left(\frac{1}{1-\delta_{1}}\right)\left(\frac{1}{1-\epsilon_1}\right)^{1-2\nu_{T}}\right|_{c_{T} k=a H},
\end{aligned}\end{equation}
with
\begin{equation}
\nu_{T}^2=\tau^2\frac{z_{T}^{\prime \prime}}{z_{T}}+\frac{1}{4},
\end{equation}
and the tensor spectral index is
\begin{equation}
n_{T}=\frac{d\ln \mathcal{P}_{T}}{d\ln k}=3-2\nu_{T}.
\end{equation}
Combing Eq.(21) and (23), we obtain the tensor to scalar ratio
\begin{equation}r \equiv \frac{\mathcal{P}_{T}}{\mathcal{P}_{\mathcal{R}}} \simeq 2^{3+2 \nu_{T}-2 \nu_{\mathcal{R}}}F \frac{c_{\mathcal{R}}^{3}}{c_{T}^{3}} \frac{\Gamma^{2}\left(\nu_{T}\right)}{\Gamma^{2}\left(\nu_{\mathcal{R}}\right)}\frac{\left(1-\epsilon_1\right)^{2\nu_T-2\nu_{\mathcal{R}}}}{\left(1-\Delta/ 2\right)^2 (1-\delta_1)}.
\end{equation}
Below we will assume that the first GB flow parameter $\delta_1$ is a constant which contains two cases: $\delta_1=0$ or $\delta_1\neq0$.

\subsection{$\delta_1=0$ }

If $\delta_1=0$, using (22),(26) we get
\begin{equation}
n_{\mathcal{R}}=4-\Big|\frac{3-\epsilon_1}{1-\epsilon_1}\Big|,
\end{equation}
\begin{equation}
r=16 \epsilon_1.
\end{equation}
The $n_{\mathcal{R}}-r$ region predicted by the model with $\epsilon_1=$ constant and $\delta_1=0$ are show in Fig.1, where the contours are the marginalized joint 68\% and 95\% confidence level regions for $n_{\mathcal{R}}$ and $r$ at the pivot scale $k_*= 0.002$ Mpc$^{-1}$ from the Planck 2018 TT,TE,EE+lowE+lensing data.
\begin{figure}\small

 \centering
  \includegraphics[width=4in]{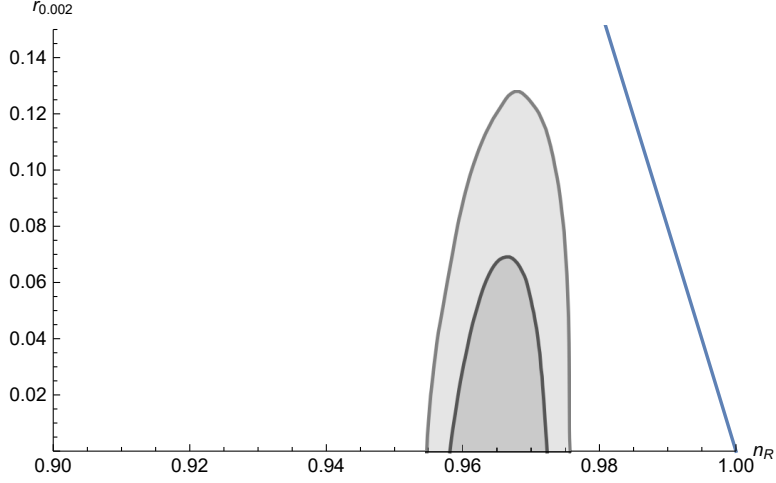}
   \caption{The $n_{\mathcal{R}}-r$ region predicted by the model with $\epsilon_1=$ constant and $\delta_1=0$ . The contours are the marginalized joint 68\% and 95\% confidence level regions for $n_{\mathcal{R}}$ and $r$ at the pivot scale $k_*= 0.002$ Mpc$^{-1}$ from the Planck 2018 TT,TE,EE+lowE+lensing data\cite{ref2}.}
   \label{fig-sp}
\end{figure}
We could see that the curve is ruled out by the observations, so we didn't interest in this case.

\subsection{$\delta_1\neq0$ }

If $\delta_1=$constant but not zero, the expression of scalar spectral index is the same as in the previous case
\begin{equation}
n_{\mathcal{R}}=4-\Big|\frac{3-\epsilon_1}{1-\epsilon_1}\Big|,
\end{equation}
and the tensor-to-scalar ratio is more complex
\begin{equation}
r=\frac{\left(16\left(\left(2 \delta_{1}^{2}-6 \delta_{1}+4\right) \epsilon_{1}+5 \delta_{1}^{2}-2 \delta_{1}\right)\right)}{\left(2-3 \delta_{1}\right)^{2}}\left(\frac{\left(3 \delta_{1}-2\right)\left(\left(\delta_{1}^{2}+2 \delta_{1}-2\right) \epsilon_{1}-2 \delta_{1}^{2}+\delta_{1}\right)}{\left(1-\delta_{1} \epsilon_{1}\right)\left(\left(2 \delta_{1}^{2}-6 \delta_{1}+4\right) \epsilon_{1}+5 \delta_{1}^{2}-2 \delta_{1}\right)}\right)^{3 / 2}
\end{equation}
Combine these expressions with the observational constraints from Planck 2018, one can obtain the  constraints on $\epsilon_1$ and $\delta_1$.
However, for a significative inflation we have $\ddot{a}>0$, which is equivalent to $\epsilon_1<1$, so only the parameter region satisfies $\epsilon_1<1$ is reasonable.
We then obtain two regions of parameter space, and show them in Fig.2.


\begin{figure}
 \begin{minipage}[t]{0.49\linewidth} 
	\centering
	\includegraphics[width=.99\textwidth]{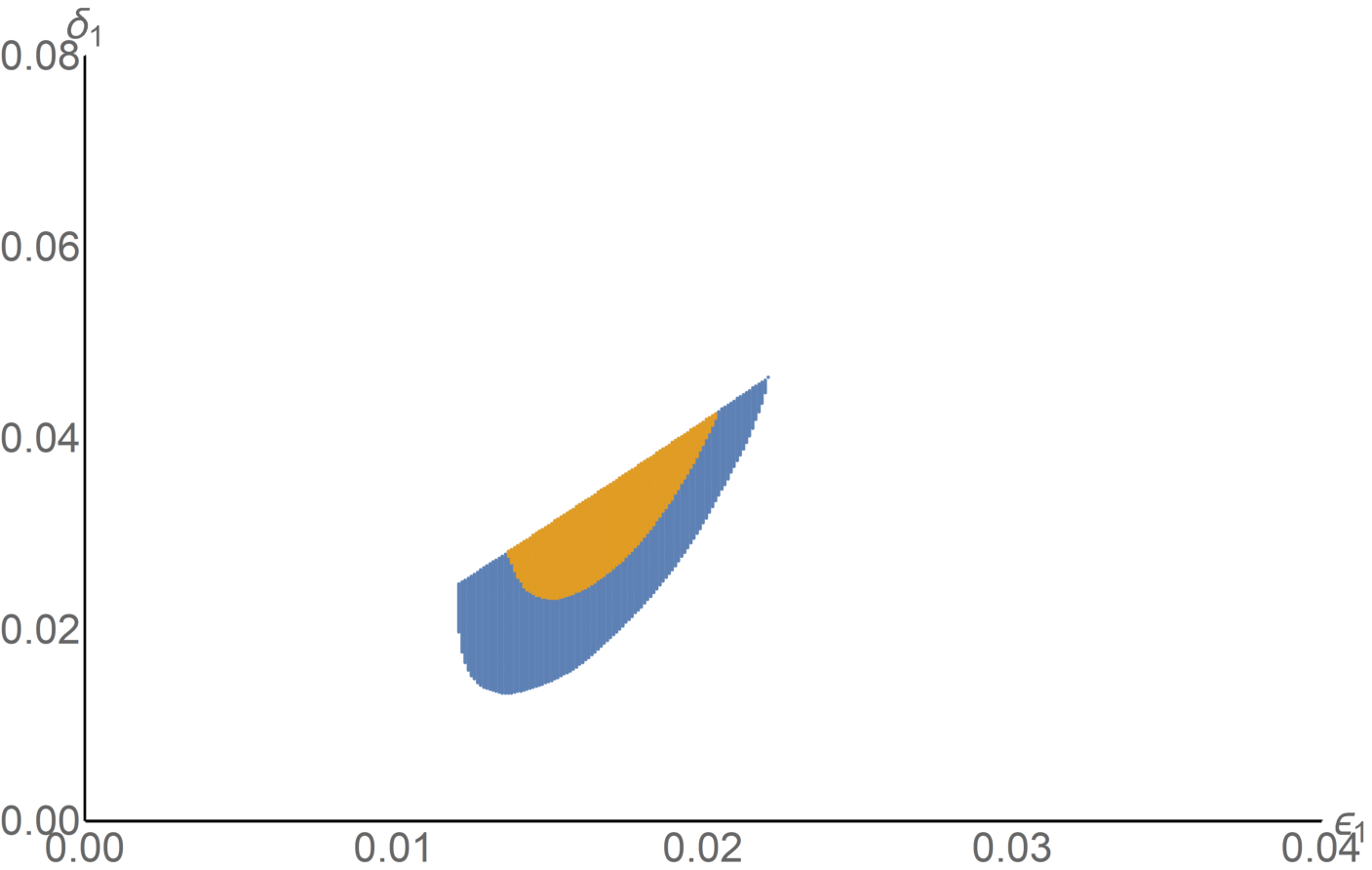}
\centerline{Region I}
	\label{fig:a} %
 \end{minipage}
 \begin{minipage}[t]{0.49\linewidth} 
	\centering
	\includegraphics[width=.99\textwidth]{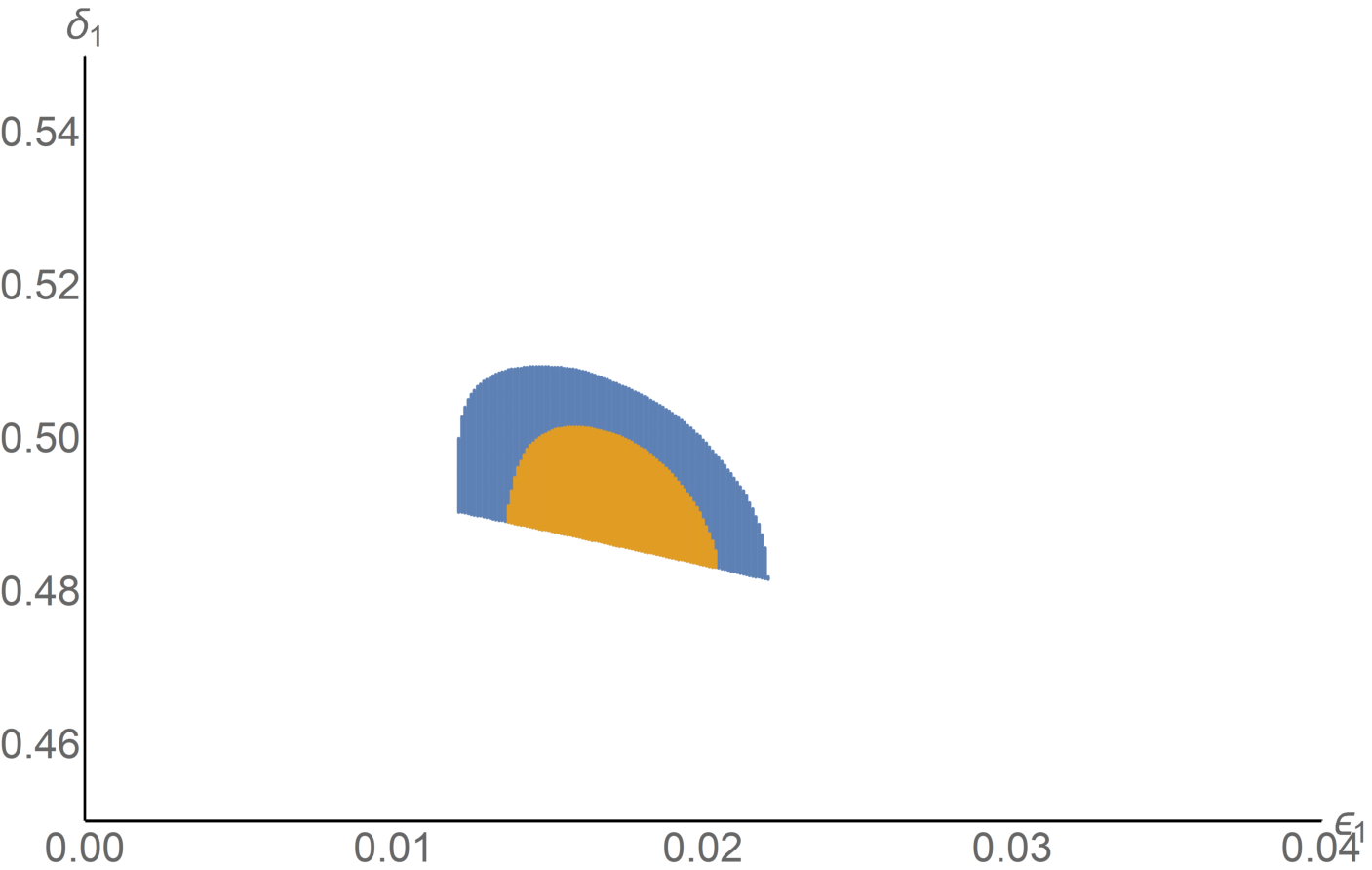}
\centerline{Region II}
	\label{fig:b}
 \end{minipage}
 \caption{The observational constraints on $\epsilon_1$ and $\delta_1$. The orange and blue regions correspond to the parameters satisfied $1\sigma$ and $2\sigma$ confidence level, respectively. }
\label{fig:1}
\end{figure}
In Fig.3, we show the $n_{\mathcal{R}}-r$ region predicted by the model for the parameters in region I of Fig.2.
\begin{figure}
 \begin{minipage}[t]{0.49\linewidth} 
	\centering
	\includegraphics[width=.99\textwidth]{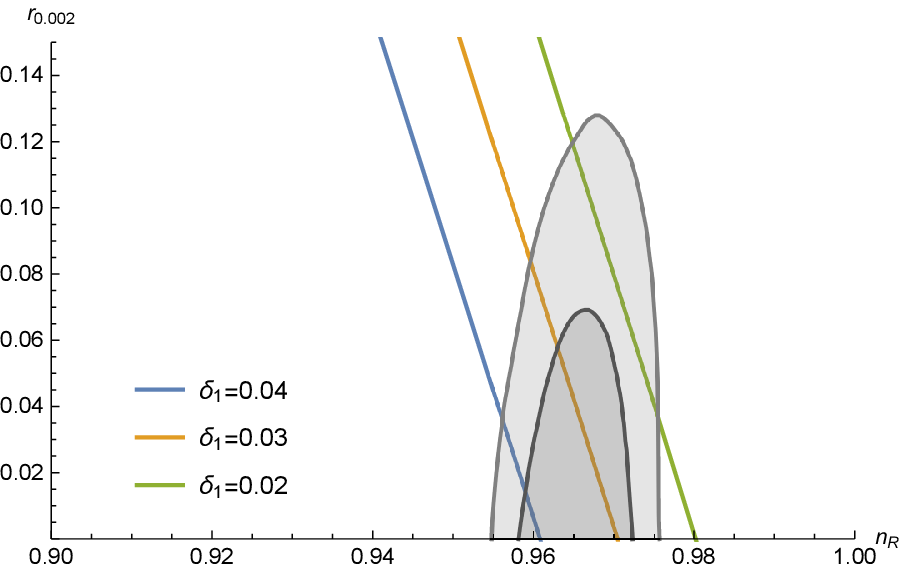}
	\label{fig:a} %
 \end{minipage}
 \begin{minipage}[t]{0.49\linewidth} 
	\centering
	\includegraphics[width=.99\textwidth]{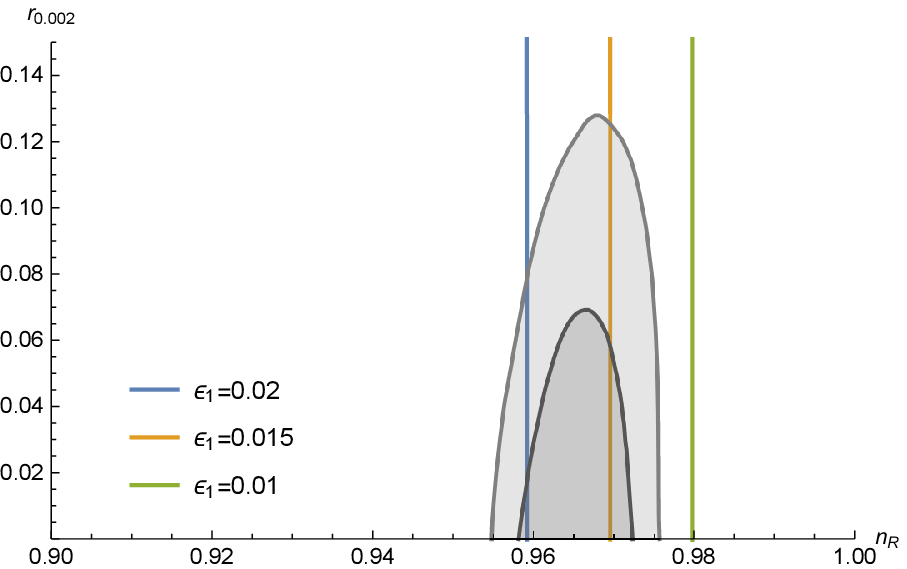}
	\label{fig:b}
 \end{minipage}
 \caption{The $n_{\mathcal{R}}-r$ region predicted by the model for the parameter region I of Fig.2.
  }
\label{fig:1}
\end{figure}
Left panel: The parameter $\delta_1$ is taken as $\delta_1=0.04,0.03,0.02$ from left to right, and as $\epsilon_1$ increase, the $n_{\mathcal{R}}-r$ dots go along the curves to the left. Right panel: The parameter $\epsilon_1$ is taken as $\epsilon_1=0.02,0.015,0.01$ from left to right, and as $\delta_1$ increase, the $n_{\mathcal{R}}-r$ dots go along the curves from top to bottom.
The $n_{\mathcal{R}}-r$ region for the parameters in region II of Fig.2 are shown In Fig.4.
\begin{figure}
 \begin{minipage}[t]{0.49\linewidth} 
	\centering
	\includegraphics[width=.99\textwidth]{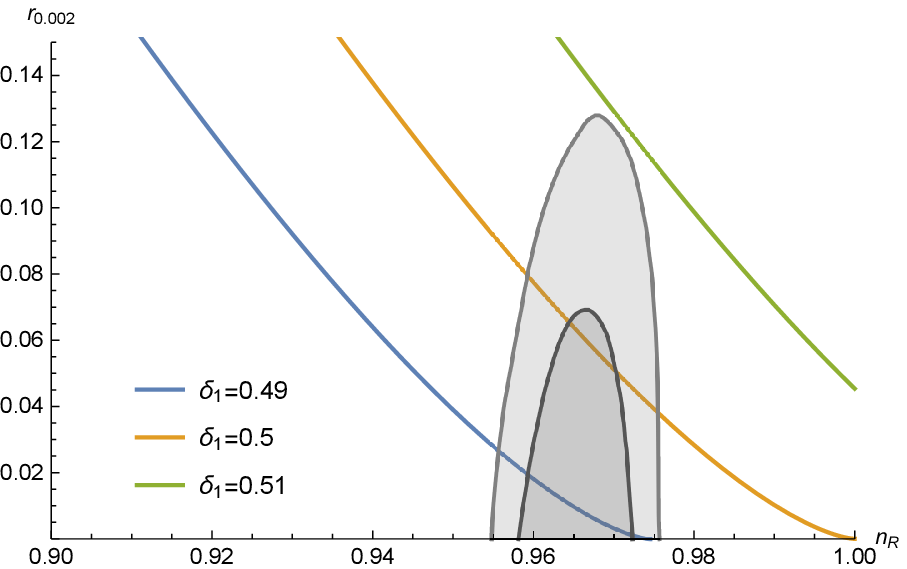}
	\label{fig:a} %
 \end{minipage}
 \begin{minipage}[t]{0.49\linewidth} 
	\centering
	\includegraphics[width=.99\textwidth]{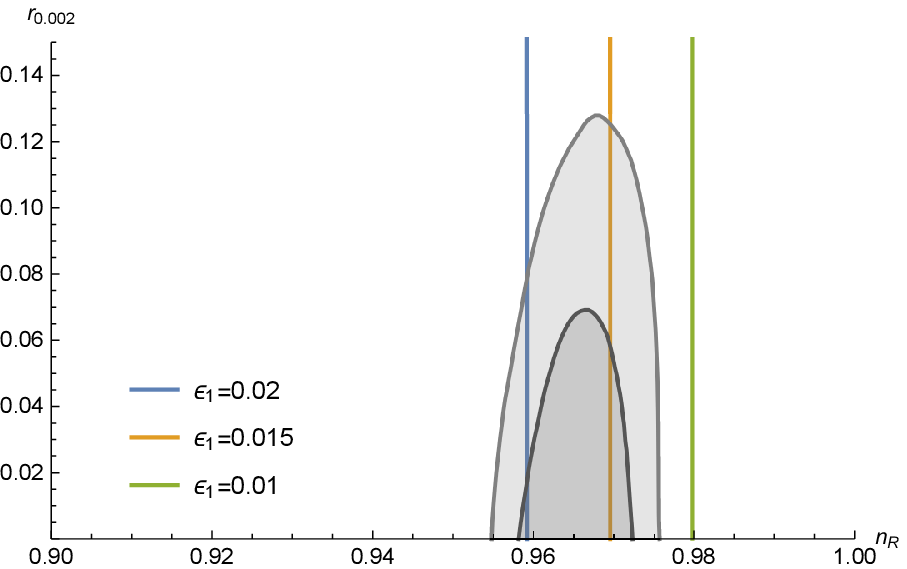}
	\label{fig:b}
 \end{minipage}
 \caption{ The $n_{\mathcal{R}}-r$ region predicted by the model for the parameter region II of Fig.2. }
\label{fig:1}
\end{figure}
Left panel:  The parameter $\delta_1$ is taken as $\delta_1=0.49,0.5,0.51$ from left to right, and as $\epsilon_1$ increase, the $n_{\mathcal{R}}-r$ dots go along the curves to the left. Right panel: The parameter $\epsilon_1$ is taken as $\epsilon_1=0.02,0.015,0.01$ from left to right. We could see that this panel is almost the same as the one in Fig.3, that is because the scalar spectral index depends only on $\epsilon_1$, so for the same chosen of $\epsilon_1$, the curves look the same. However, different from that panel, in this time the $n_{\mathcal{R}}-r$ dots go along the curves from bottom to top as $\delta_1$ increase.

We could see that for the choices  of parameter space in Fig.2, the inflation prediction is in agreement with the observational constraints. However, the constant-roll inflation with $\epsilon_1=$ constant is no natural end, hence an additional mechanism is required to stop it.

\section{Constant-roll inflation with constant $\epsilon_2$  \label{sec3}}
In this section, we assume that the second horizon flow parameter $\epsilon_2$ is a constant.  
So from the relations (17) and using the definition of flow parameter (6), 
we obtain that to the first order approximation of $\epsilon_1$, the relation between $aH$ and $\tau$ is\cite{ref16,ref18}
\begin{eqnarray}
&&aH\simeq -\frac{1}{\tau }\left(1+\frac{\epsilon_1}{1-\epsilon_2}\right),
\label{infp9}
\end{eqnarray}
then, we obtain the power spectrum of the scalar perturbation and tensor perturbation
\begin{equation}\begin{aligned}
\mathcal{P}_{\mathcal{R}} &\left.\simeq \frac{2^{2 \nu_{\mathcal{R}}-3} c_{\mathcal{R}}^{-3}}{F} \frac{H^{2}}{4 \pi^{2}} \frac{\Gamma^{2}\left(\nu_{\mathcal{R}}\right)}{\Gamma^{2}(3 / 2)}\left(1-\frac{\Delta}{2}\right)^2\left(1+\frac{\epsilon_1}{1-\epsilon_2}\right)^{1-2\nu_{\mathcal{R}}}\right|_{c_{\mathcal{R}} k=a H},
\end{aligned}\end{equation}
\begin{equation}\begin{aligned}
\mathcal{P}_{T} &\left.\simeq 2^{2 \nu_{T}} c_{T}^{-3} \frac{H^{2}}{4 \pi^{2}} \frac{\Gamma^{2}\left(\nu_{T}\right)}{\Gamma^{2}(3 / 2)}\left(\frac{1}{1-\delta_{1}}\right)\left(1+\frac{\epsilon_1}{1-\epsilon_2}\right)^{1-2\nu_{T}}\right|_{c_{T} k=a H},
\end{aligned}\end{equation}
with the scalar spectral index and the tensor-to-scalar ratio are
\begin{equation}
n_{\mathcal{R}}-1=\frac{d\ln \mathcal{P}_{\mathcal{R}}}{d\ln k}=3-2\nu_{\mathcal{R}},
\end{equation}
\begin{equation}
r \equiv \frac{\mathcal{P}_{T}}{\mathcal{P}_{\mathcal{R}}} \simeq 2^{3+2 \nu_{T}-2 \nu_{\mathcal{R}}}F \frac{c_{\mathcal{R}}^{3}}{c_{T}^{3}} \frac{\Gamma^{2}\left(\nu_{T}\right)}{\Gamma^{2}\left(\nu_{\mathcal{R}}\right)}\frac{\left(1+\frac{\epsilon_1}{1-\epsilon_2}\right)^{2 \nu_{\mathcal{R}}-2 \nu_{T}}}{\left(1-\Delta/ 2\right)^2 (1-\delta_1)}.
\end{equation}

In the following, we still assume that the first GB flow parameter $\delta_1$ is a constant, which contains two cases: $\delta_1=0$ or $\delta\neq0$.

\subsection{$\delta_1=0$ }
If $\delta_1=0$, the model recovers to the case without GB coupling, and such case have been discussed in other works\cite{ref16}. Here we list some useful results.

To the first order approximation of $\epsilon_1$, the scalar spectral index can be approximated as
\begin{equation}
n_{\mathcal{R}}\simeq4-|\epsilon_2+3|+\frac{\left(2 \epsilon_2^2+7 \epsilon_2+6\right)\epsilon_1}{(\epsilon_2-1)|\epsilon_2+3|},
\end{equation}
and the tensor-to-scalar ratio
\begin{equation}
r \simeq 2^{3-\left|3+\epsilon_{2}\right|}\left(\frac{\Gamma[3 / 2]}{\Gamma\left[\left|3+\epsilon_{2}\right| / 2\right]}\right)^{2} 16 \epsilon_{1}.
\end{equation}
Since $\epsilon_{2}$ is a constant, we can integer the definition (6) and  get the relation between $\epsilon_1$ and $\epsilon_2$
\begin{equation}
\epsilon_{1}(N)=\exp \left(-\epsilon_{2} N\right),
\end{equation}
where $N$ is the e-folding number before the end of inflation, and we have use the condition $\epsilon_1 (N=0)=1$ at the end of inflation.
The $n_{\mathcal{R}}-r$ predictions compare with the Planck 2018 data are show in Fig.5(blue line) with $N=60$. We can see that the result is ruled out by the observations. So it is always to assume that the second Hubble flow parameter $\eta_H$ is a constant in the constant-roll inflationary models, we will discuss this case in the next section.
\begin{figure}\small

 \centering
  \includegraphics[width=4in]{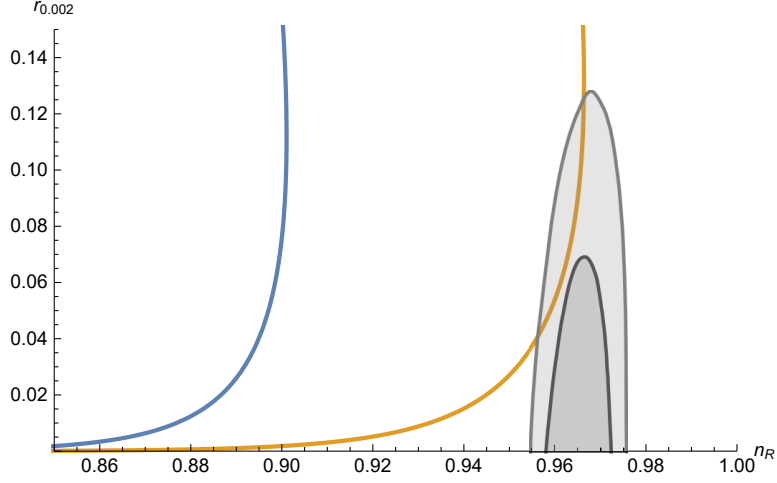}
   \caption{The $n_{\mathcal{R}}-r$ region predicted by the model with $\delta_1=0$ and  constant $\epsilon_1$(blue line) or constant $\eta_H$(orange line).}
   \label{fig-sp}
\end{figure}

\subsection{$\delta_1\neq0$ }

If $\delta_1=$constant but not zero, to the first order of $\epsilon_1$, the expression of scalar spectral index is
\begin{equation}
n_{\mathcal{R}}\simeq1-\frac{2\left(3+\epsilon_{2}\right)\left(-2 \epsilon_{2}+2 \epsilon_{2}^{2}+\delta_{1}\left(2+3 \epsilon_{2}-3 \epsilon_{2}^{2}\right)+\delta_{1}^{2}\left(-5-\epsilon_{2}+\epsilon_{2}^{2}\right)\right)}{3\left(2 \delta_{1}-5 \delta_{1}^{2}\right)\left(1-\epsilon_{2}\right)} \epsilon_{1},
\end{equation}
and the tensor-to-scalar ratio
\begin{equation}\begin{aligned}
r\simeq &-\frac{16}{3\left(3 \delta_{1}-2\right)\left(5 \delta_{1}-2\right)} \sqrt{\frac{2-7\delta_{1}+6 \delta_{1}^2}{2-5 \delta_{1}}}\Big(30 \delta_{1}^{3}-27 \delta_{1}^{2}+6 \delta_{1} \\
&+\epsilon_{1}\left(45 \delta_{1}^{4}-69 \delta_{1}^{3}-6 \delta_{1}^{2}+42 \delta_{1}-12+\left(-2 \delta_{1}^{3}+7 \delta_{1}^{2}-7 \delta_{1}+2\right) \epsilon_{2}\left(\epsilon_{2}+3\right) \ln 4\right. \\
&\left.+2\left(2 \delta_{1}^{3}-7 \delta_{1}^{2}+7 \delta_{1}-2\right) \epsilon_{2}\left(\epsilon_{2}+3\right)(\gamma-2+\ln 4)\right)\Big),
\end{aligned}\end{equation}

with $\gamma\approx0.577216$ is the Euler-Mascheroni constant.
Combine these expressions with the observational constraints and adopting the relation $\epsilon_{1}(N)=\exp \left(-\epsilon_{2} N\right)$, one can obtain the constraints on $\epsilon_1$ and $\delta_1$.
By setting the e-folding number $N=60$, we find three regions of parameter space, and show them in Fig.6.
\begin{figure}
 \begin{minipage}[t]{0.32\linewidth} 
	\centering
	\includegraphics[width=.99\textwidth]{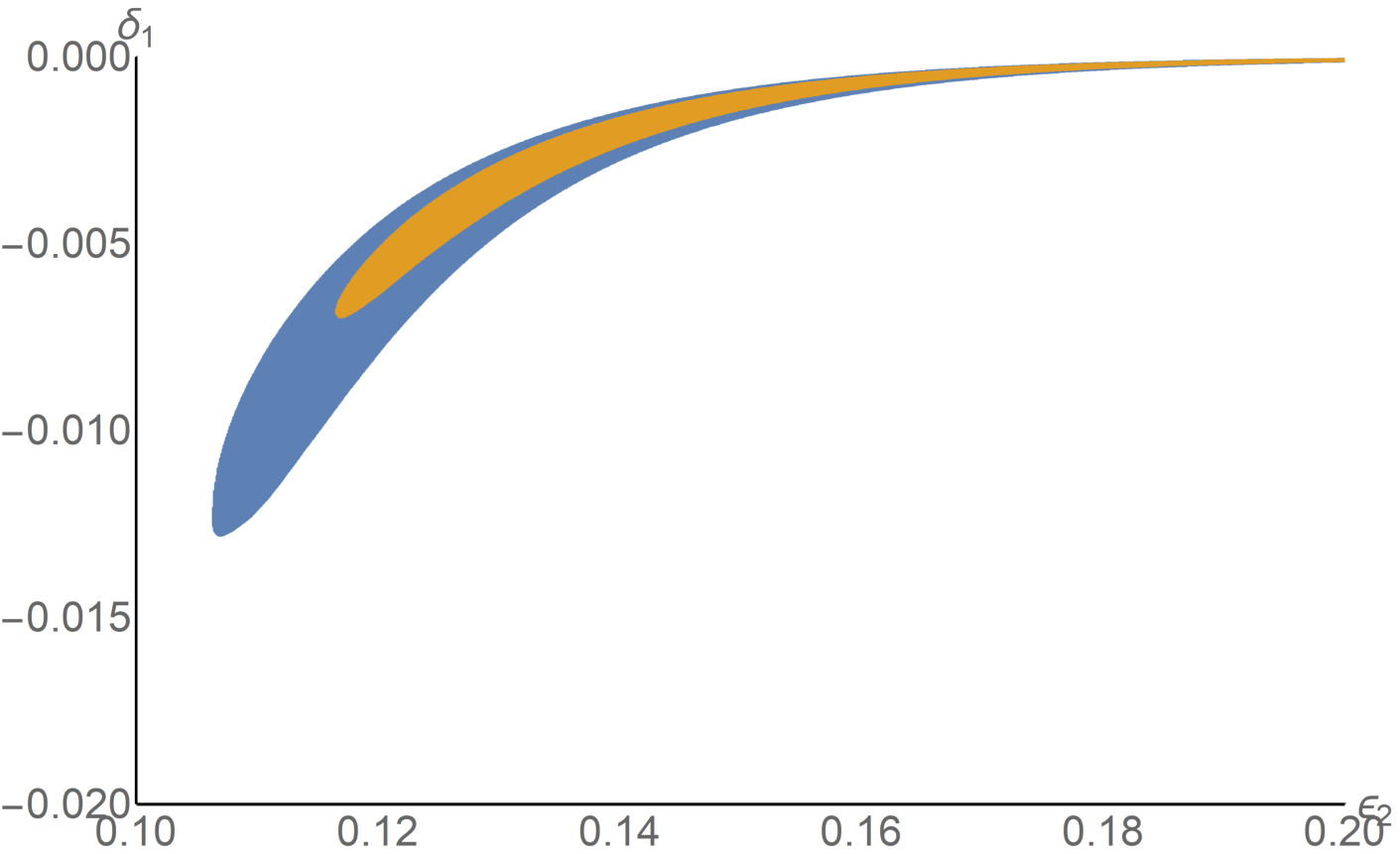}
\centerline{Region I}
	\label{fig:a} %
 \end{minipage}
 \begin{minipage}[t]{0.32\linewidth} 
	\centering
	\includegraphics[width=.99\textwidth]{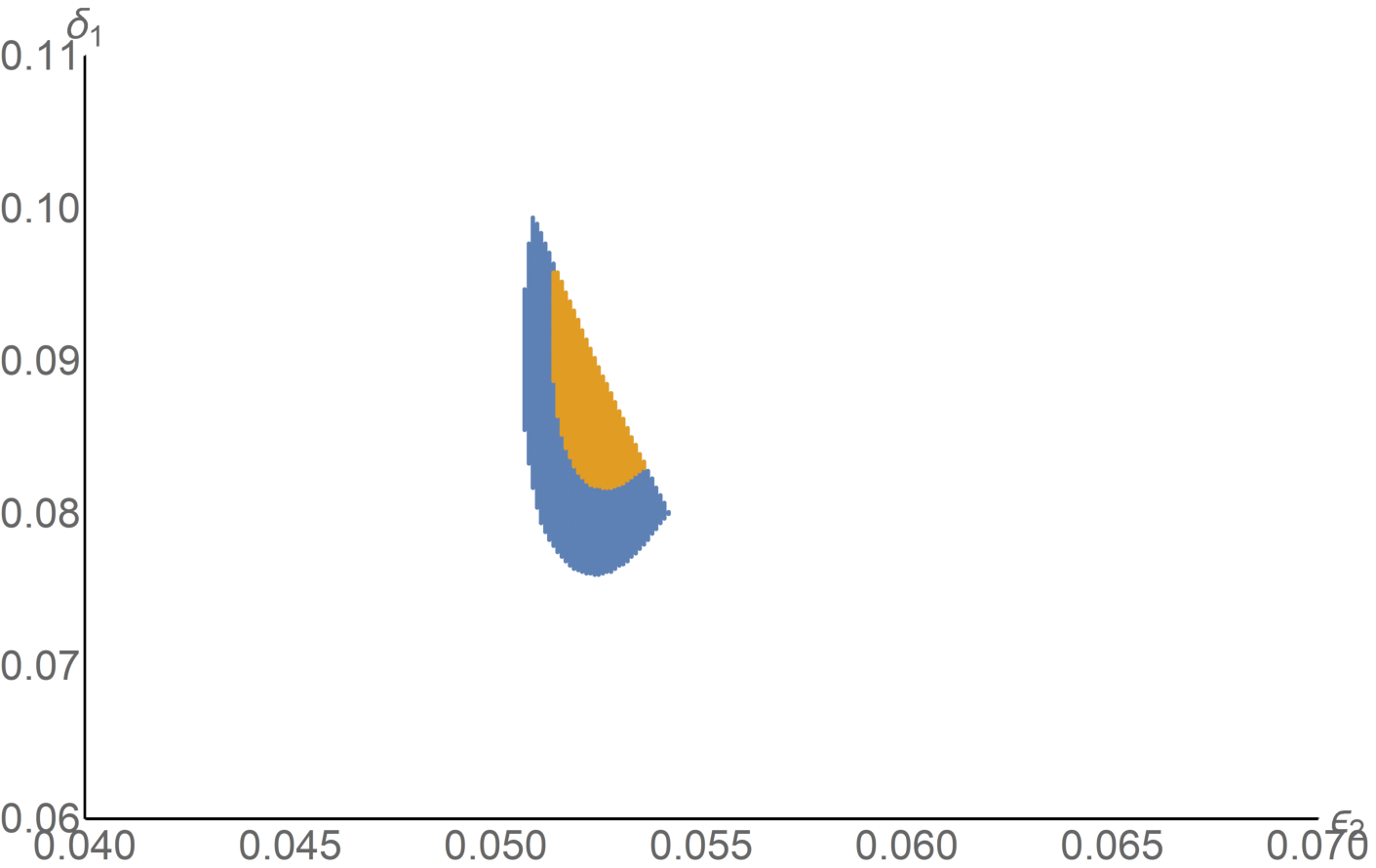}
\centerline{Region II}
	\label{fig:b}
 \end{minipage}
 \begin{minipage}[t]{0.32\linewidth} 
	\centering
	\includegraphics[width=.99\textwidth]{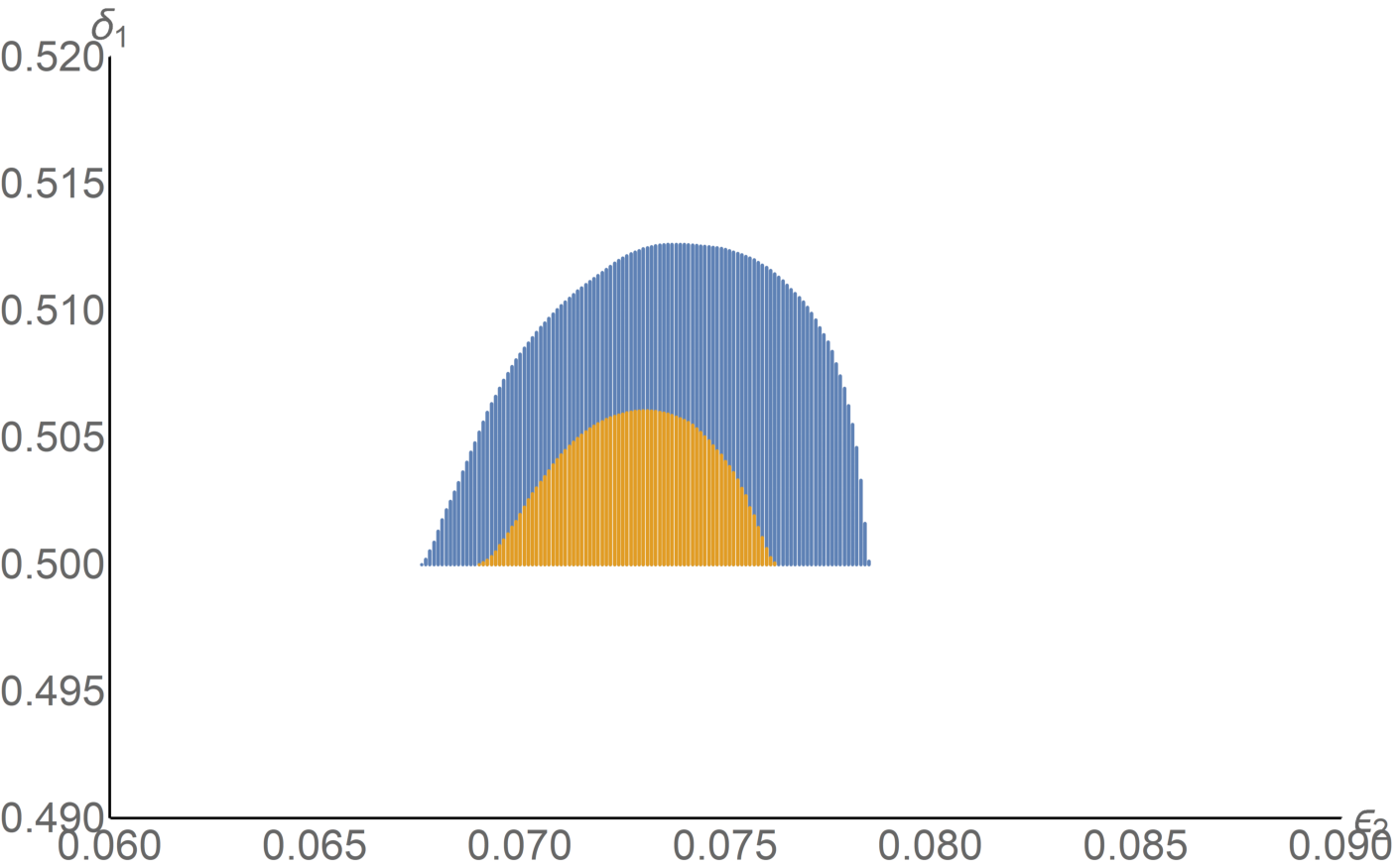}
\centerline{Region III}
	\label{fig:b}
 \end{minipage}
 \caption{The observational constraints on $\epsilon_2$ and $\delta_1$ with the e-folding number $N=60$. The orange and blue regions correspond to the parameters satisfied $1\sigma$ and $2\sigma$ confidence level, respectively. }
\label{fig:1}
\end{figure}

The observational constraint on $n_{\mathcal{R}}-r$ for the parameters in region I of Fig.6 are show in Fig.7.
\begin{figure}
 \begin{minipage}[t]{0.49\linewidth} 
	\centering
	\includegraphics[width=.99\textwidth]{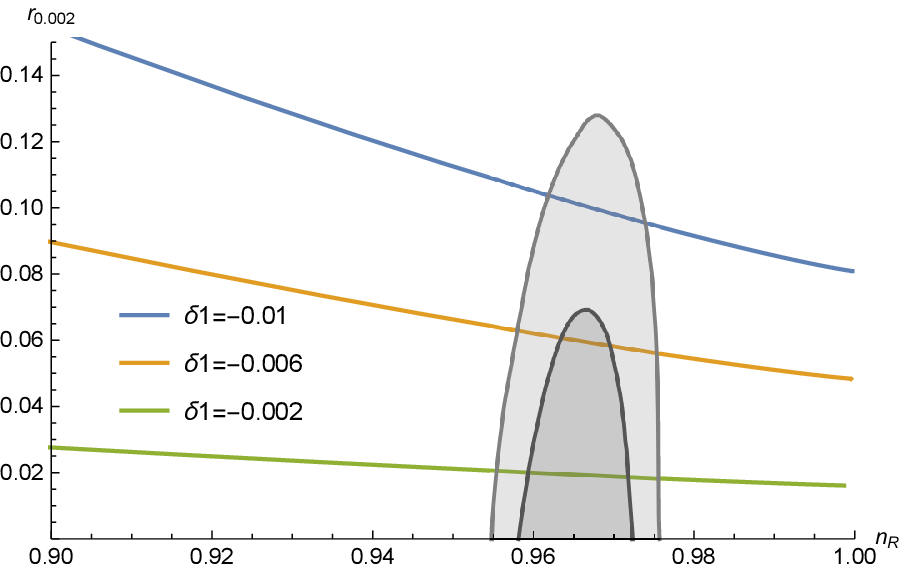}
	\label{fig:a} %
 \end{minipage}
 \begin{minipage}[t]{0.49\linewidth} 
	\centering
	\includegraphics[width=.99\textwidth]{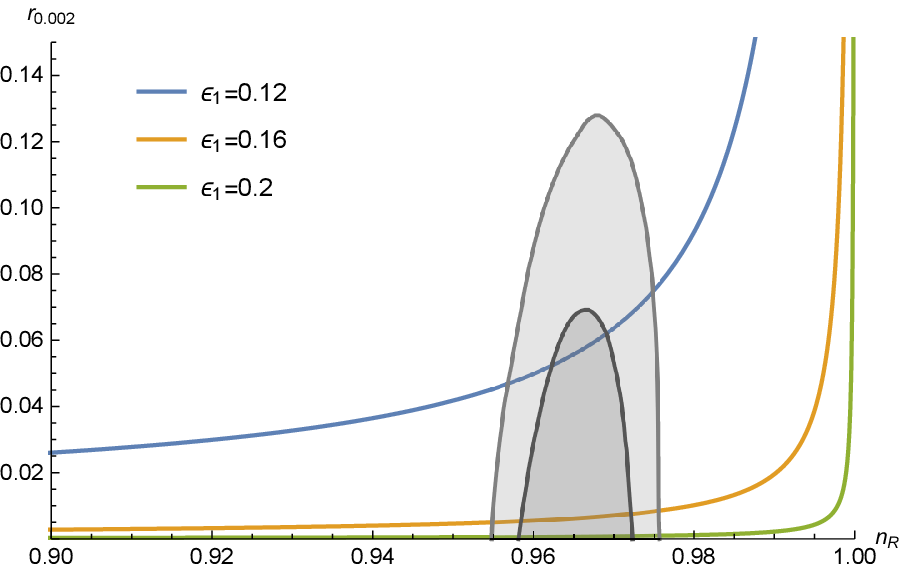}
	\label{fig:b}
 \end{minipage}
 \caption{ The $n_{\mathcal{R}}-r$ region predicted by the model for the parameter region I of Fig.6. }
\label{fig:1}
\end{figure}
Left panel: The parameter $\delta_1$ is taken as $\delta_1=-0.01,-0.006,-0.002$ from top to bottom, and as $\epsilon_1$ increase, the $n_{\mathcal{R}}-r$ dots go along the curves from left to right. Right panel: The parameter $\epsilon_1$ is taken as $\epsilon_1=0.12,0.16,0.2$ from left to right, and as $\delta_1$ increase, the $n_{\mathcal{R}}-r$ dots go along the curves to the  left.

Similarly, in Fig.8, we show the $n_{\mathcal{R}}-r$ region predicted for the parameters in region II of Fig.6.
\begin{figure}
 \begin{minipage}[t]{0.49\linewidth} 
	\centering
	\includegraphics[width=.99\textwidth]{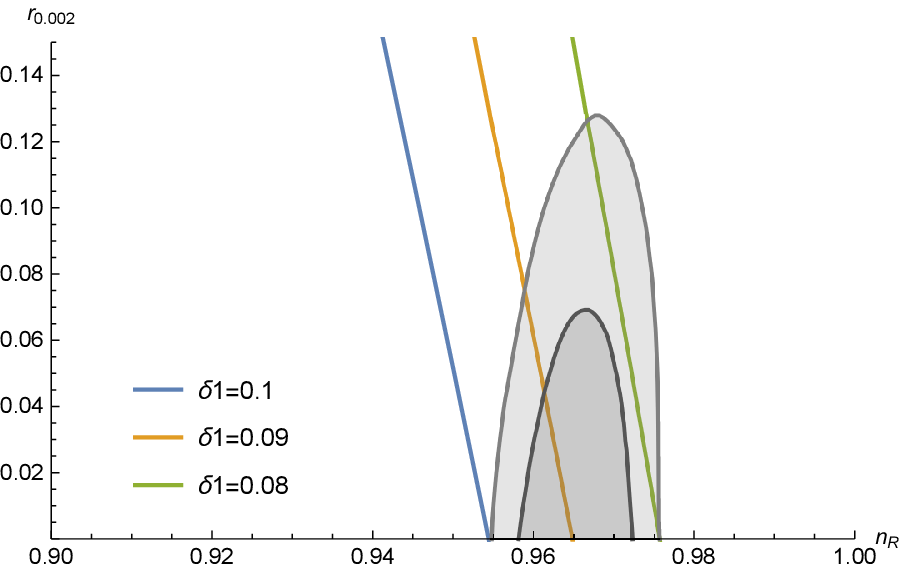}
	\label{fig:a} %
 \end{minipage}
 \begin{minipage}[t]{0.49\linewidth} 
	\centering
	\includegraphics[width=.99\textwidth]{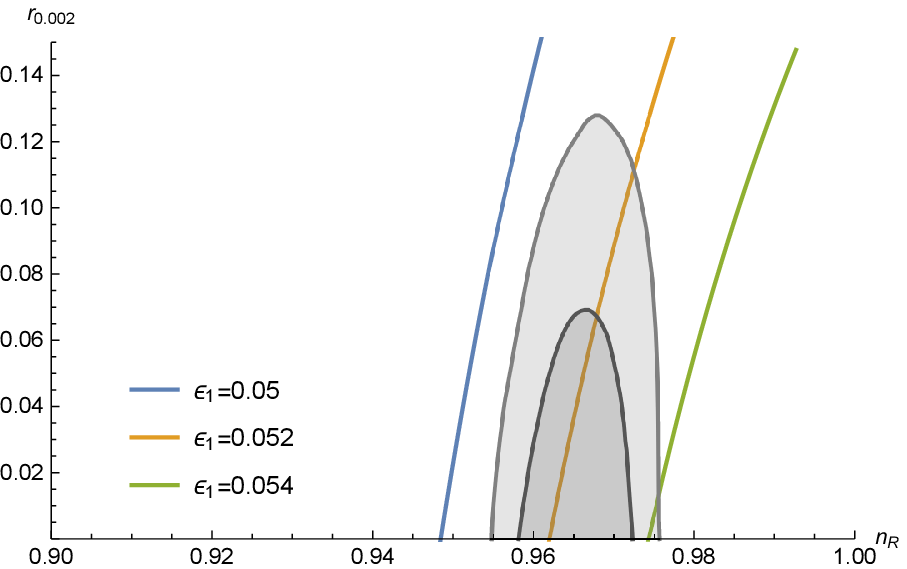}
	\label{fig:b}
 \end{minipage}
 \caption{ The $n_{\mathcal{R}}-r$ region predicted by the model for the parameter region II of Fig.6. }
\label{fig:1}
\end{figure}
Left panel: The parameter $\delta_1$ is taken as $\delta_1=0.1,0.09,0.08$ from left to right, and as $\epsilon_1$ increase, the $n_{\mathcal{R}}-r$ dots go along the curves from top to bottom. Right panel: $\epsilon_1$ is taken as $\epsilon_1=0.05,0.052,0.054$ from left to right, and as $\delta_1$ increase, the dots go along the curves from top to bottom.

Finally, the $n_{\mathcal{R}}-r$ region for the parameters in region III of Fig.6 are show in Fig.9.
\begin{figure}
 \begin{minipage}[t]{0.49\linewidth} 
	\centering
	\includegraphics[width=.99\textwidth]{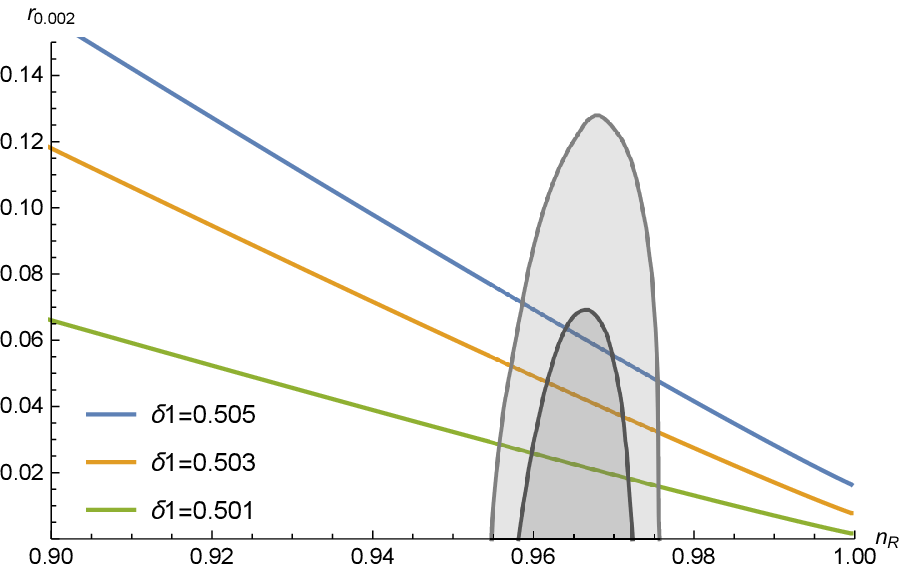}
	\label{fig:a} %
 \end{minipage}
 \begin{minipage}[t]{0.49\linewidth} 
	\centering
	\includegraphics[width=.99\textwidth]{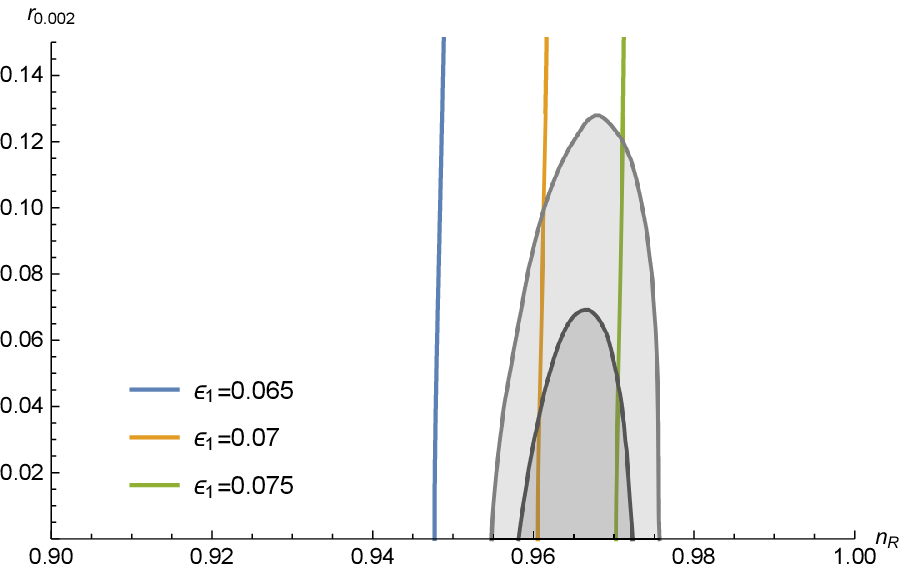}
	\label{fig:b}
 \end{minipage}
 \caption{ The $n_{\mathcal{R}}-r$ region predicted by the model for the parameter region III of Fig.6. }
\label{fig:1}
\end{figure}
With  $\delta_1$ is taken as $\delta_1=0.505,0.503,0.501$ from left to right in the left panel, and as $\epsilon_1$ increase, the $n_{\mathcal{R}}-r$ dots go along the curves to the right. In the right panel, $\epsilon_1$ is taken as $\epsilon_1=0.065,0.07,0.075$ from left to right,  and the dots go along the curves from bottom to top as $\delta_1$ increase.

\section{Constant-roll inflation with constant $\eta_H$  \label{sec3}}
In this section, the second Hubble flow parameter $\eta_{H}=-\ddot{H}/(2H\dot{H})$ is assumed to be a constant. Combine with the Eq.(17), the relation between $aH$ and $\tau$ to the first order approximation of $\epsilon_1$ can be obtained as\cite{ref18}
\begin{equation}
a H \approx-\frac{1}{\tau}\left(1+\frac{\epsilon_{1}}{1+2 \eta_{H}}\right).
\end{equation}
Substitute this relation to the expressions of power spectrum, we get
\begin{equation}\begin{aligned}
\mathcal{P}_{\mathcal{R}} &\left.\simeq \frac{2^{2 \nu_{\mathcal{R}}-3} c_{\mathcal{R}}^{-3}}{F} \frac{H^{2}}{4 \pi^{2}} \frac{\Gamma^{2}\left(\nu_{\mathcal{R}}\right)}{\Gamma^{2}(3 / 2)}\left(1-\frac{\Delta}{2}\right)^2\left(1+\frac{\epsilon_{1}}{1+2 \eta_{H}}\right)^{1-2\nu_{\mathcal{R}}}\right|_{c_{\mathcal{R}} k=a H},
\end{aligned}\end{equation}
\begin{equation}\begin{aligned}
\mathcal{P}_{T} &\left.\simeq 2^{2 \nu_{T}} c_{T}^{-3} \frac{H^{2}}{4 \pi^{2}} \frac{\Gamma^{2}\left(\nu_{T}\right)}{\Gamma^{2}(3 / 2)}\left(\frac{1}{1-\delta_{1}}\right)\left(1+\frac{\epsilon_{1}}{1+2 \eta_{H}}\right)^{1-2\nu_{T}}\right|_{c_{T} k=a H},
\end{aligned}\end{equation}
with the scalar spectral index and the tensor-to-scalar ratio are
\begin{equation}
n_{\mathcal{R}}-1=3-2\nu_{\mathcal{R}},
\end{equation}
\begin{equation}
r \equiv \frac{\mathcal{P}_{T}}{\mathcal{P}_{\mathcal{R}}} \simeq 2^{3+2 \nu_{T}-2 \nu_{\mathcal{R}}}F \frac{c_{\mathcal{R}}^{3}}{c_{T}^{3}} \frac{\Gamma^{2}\left(\nu_{T}\right)}{\Gamma^{2}\left(\nu_{\mathcal{R}}\right)}\frac{\left(1+\frac{\epsilon_{1}}{1+2 \eta_{H}}\right)^{2 \nu_{\mathcal{R}}-2 \nu_{T}}}{\left(1-\Delta/ 2\right)^2 (1-\delta_1)}.
\end{equation}

Similarly as in the previous section, we set the parameter $\delta_1$ to be a constant and discuss two cases: $\delta_1=0$ or $\delta\neq0$.

\subsection{$\delta_1=0$ }
If $\delta_1=0$ and $\eta_H=constant$, which is the original constant-roll  model without GB coupling\cite{ref3}.
In this case, to the first order approximation of $\epsilon_1$, the scalar spectral index can be approximated as
\begin{equation}
n_{\mathcal{R}}\simeq4-\mid2 \eta_H -3\mid+\frac{2 \left(4 \eta_H^2+5 \eta_H -6\right) \epsilon_1}{\mid2 \eta_H -3\mid (2 \eta_H +1)},
\end{equation}
and the tensor-to-scalar ratio
\begin{equation}
r \simeq 2^{3-\left|3-2\eta_{H}\right|}\left(\frac{\Gamma[3 / 2]}{\Gamma\left[\left|3-2\eta_{H}\right| / 2\right]}\right)^{2} 16 \epsilon_{1}.
\end{equation}
Since $\eta_H$ is a constant, from the definition of flow parameters(5) and (6), and the condition $\epsilon_{1}(N=0)=1$, we obtain the relation\cite{ref18}
\begin{equation}
\epsilon_{1}(N)=\frac{\eta_H \exp \left(2 \eta_H N\right)}{\exp \left(2 \eta_H N\right)+\eta_H-1}.
\end{equation}
Then we get the $n_{\mathcal{R}}-r$ predictions  with $N=60$ and show them in Fig.5(orange line). For a constant $\eta_H$, the result is consistent with the observations at $2\sigma$ confidence level. And we updated the constraint to $\eta_H$ with the Planck 2018 data, which is $-0.0163<\eta_H<-0.0006$.

\subsection{$\delta_1\neq0$ }
If $\delta_1=$constant but not zero, the scalar spectral index and the tensor-to-scalar ratio to the first order of $\epsilon_1$ is obtained as
\begin{equation}
n_{\mathcal{R}}\simeq 1-\frac{2\left(2 \eta_{H}-3\right)\left(4 \eta_{H}+8 \eta_{H}^{2}+\delta_{1}\left(2-6 \eta_{H}-12 \eta_{H}^{2}\right)+\delta_{1}^{2}\left(-5+2 \eta_{H}+4 \eta_{H}^{2}\right)\right)}{3 \delta_{1}\left(5 \delta_{1}-2\right)\left(1+2 \eta_{H}\right)} \epsilon_{1},
\end{equation}
and
\begin{equation}\begin{aligned}
r\simeq &-\frac{16}{3\left(3 \delta_{1}-2\right)\left(5 \delta_{1}-2\right)} \sqrt{\frac{2-7\delta_{1}+6 \delta_{1}^2}{2-5 \delta_{1}}}\Big(30 \delta_{1}^{3}-27 \delta_{1}^{2}+6 \delta_{1} \\
&+\epsilon_{1}\left(45 \delta_{1}^{4}-69 \delta_{1}^{3}-6 \delta_{1}^{2}+42 \delta_{1}-12+2\left(-2 \delta_{1}^{3}+7 \delta_{1}^{2}-7 \delta_{1}+2\right) \eta_H\left(2\eta_H-3\right) \ln 4\right. \\
&\left.+4\left(2 \delta_{1}^{3}-7 \delta_{1}^{2}+7 \delta_{1}-2\right) \eta_H\left(2\eta_H-3\right)(\gamma-2+\ln 4)\right)\Big).
\end{aligned}\end{equation}

Using the relation (48) and  setting the e-folds $N=60$,  we find four reasonable regions of parameter space, which are show in Fig.10.
\begin{figure}
 \begin{minipage}[t]{0.49\linewidth} 
	\centering
	\includegraphics[width=.99\textwidth]{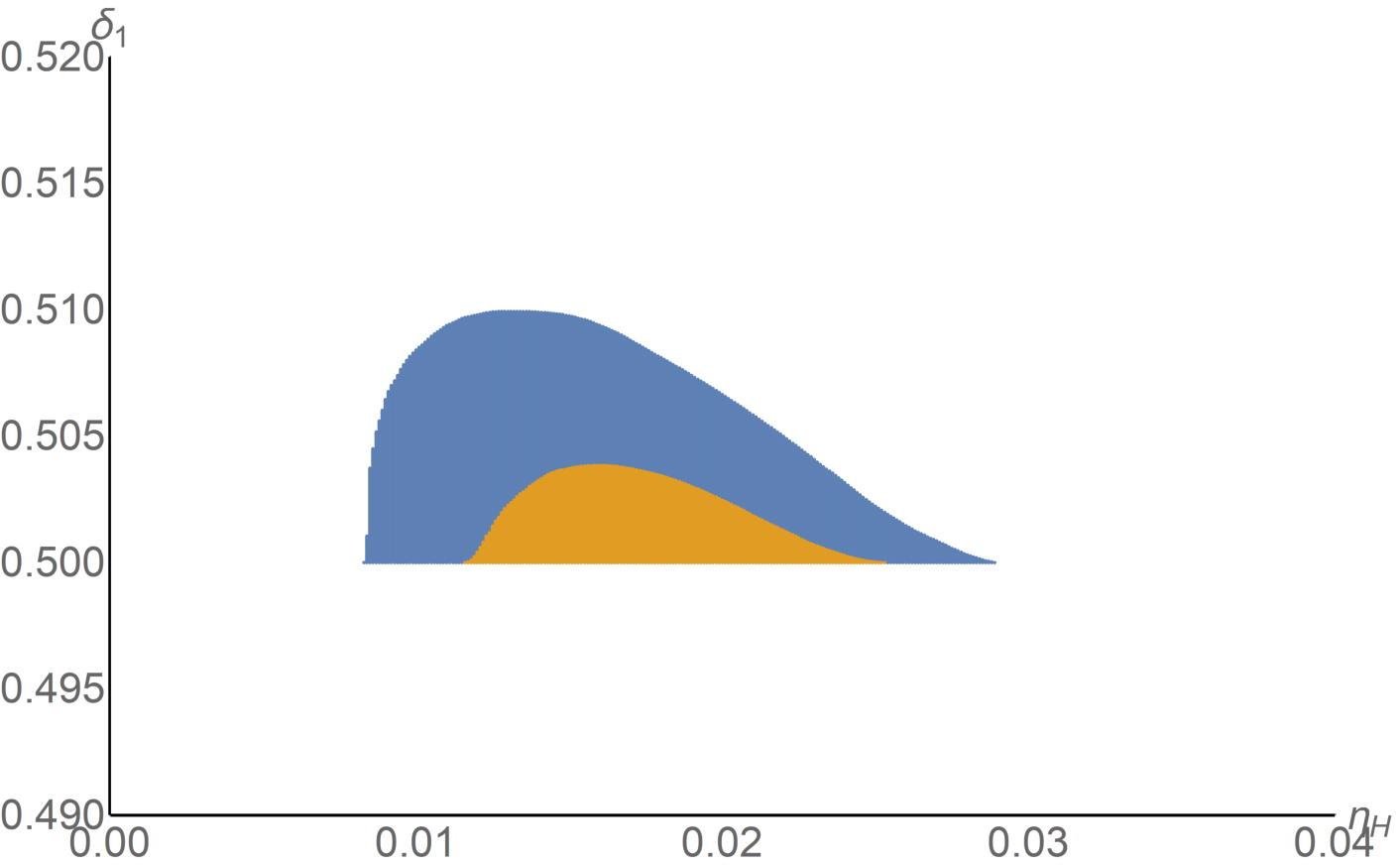}
\centerline{Region I}
	\label{fig:a} %
 \end{minipage}
 \begin{minipage}[t]{0.49\linewidth} 
	\centering
	\includegraphics[width=.99\textwidth]{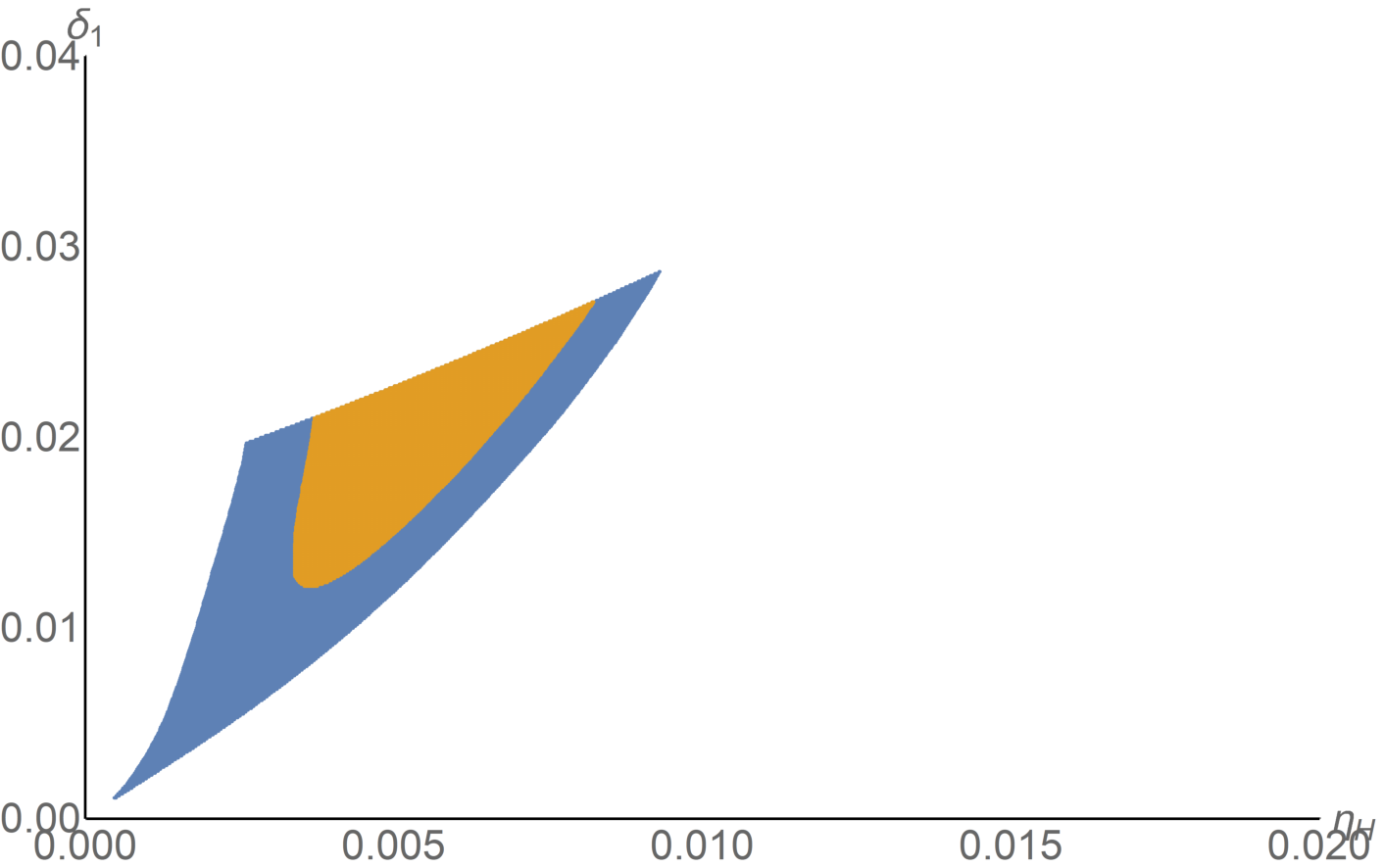}
\centerline{Region II}
	\label{fig:b}
 \end{minipage}
 \begin{minipage}[t]{0.49\linewidth} 
	\centering
	\includegraphics[width=.99\textwidth]{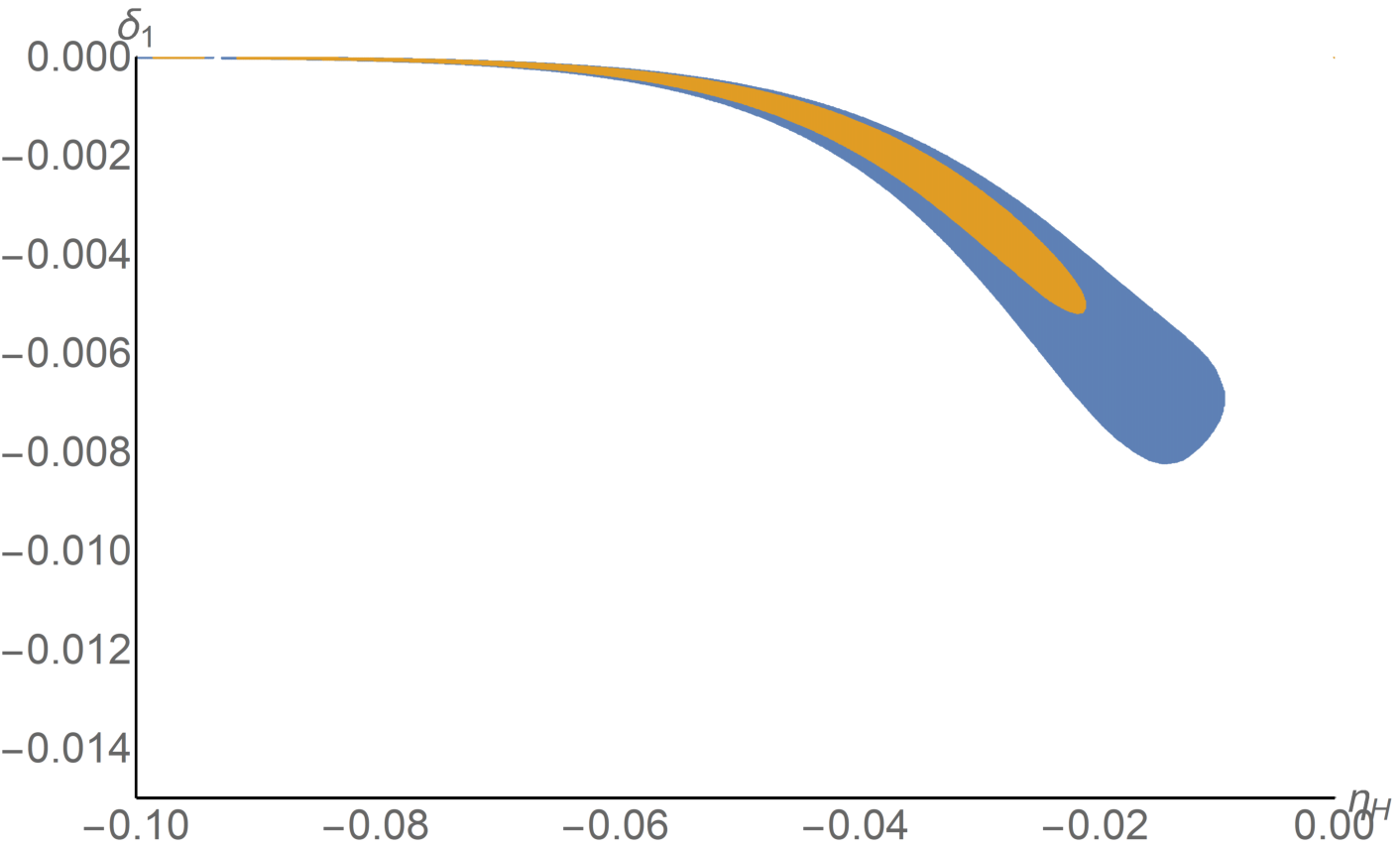}
\centerline{Region III}
	\label{fig:b}
 \end{minipage}
 \begin{minipage}[t]{0.49\linewidth} 
	\centering
	\includegraphics[width=.99\textwidth]{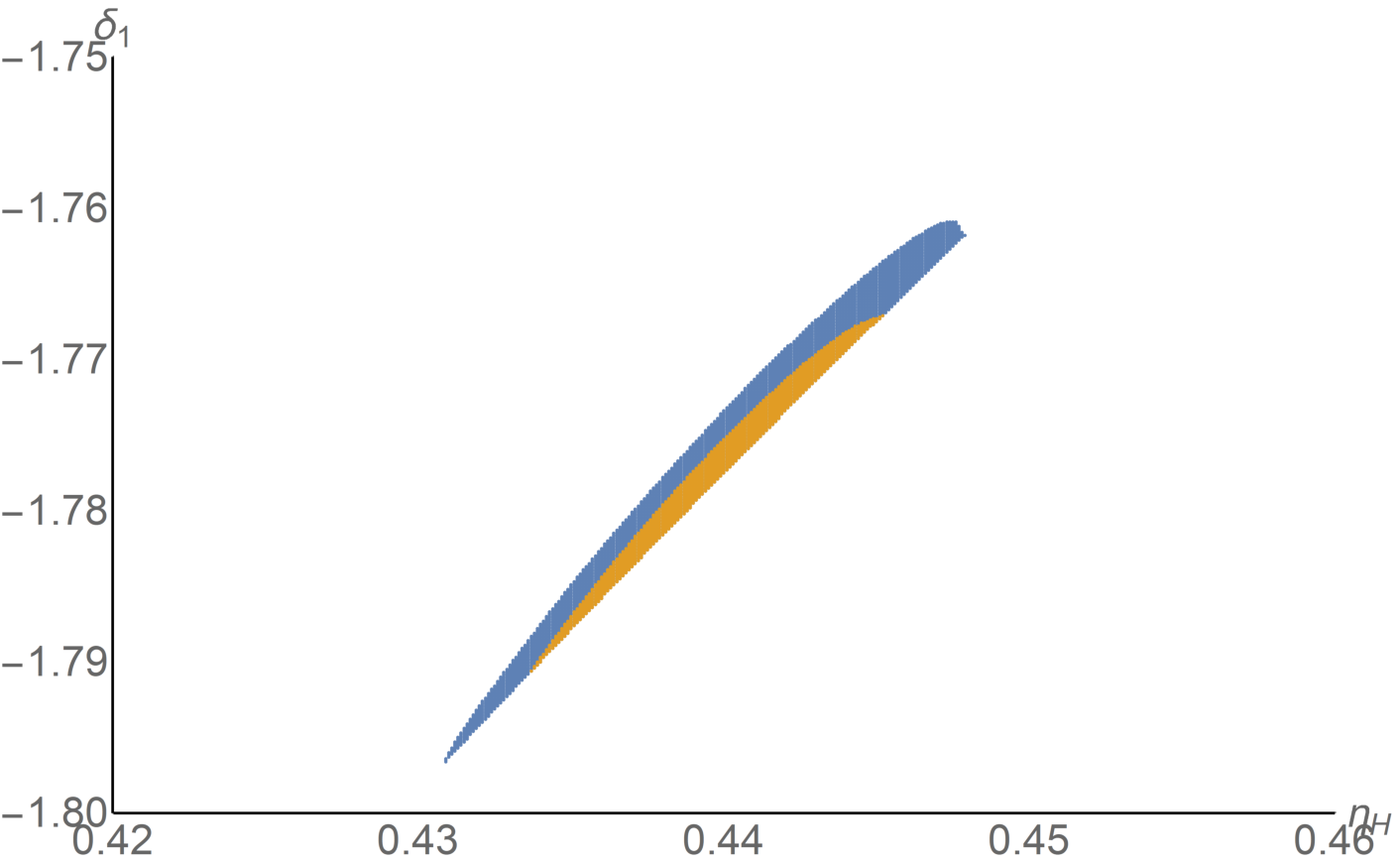}
\centerline{Region IV}
	\label{fig:b}
 \end{minipage}
 \caption{The observational constraints on $\eta_H$ and $\delta_1$ with the e-folding number $N=60$. The orange and blue regions correspond to the parameters satisfied $1\sigma$ and $2\sigma$ confidence level, respectively. }
\label{fig:1}
\end{figure}
We can see that in this case of constant-roll inflation, the absolute of GB flow parameter $\delta_1$ can be larger then one(region IV), which is different from the conventional slow-roll inflation with GB coupling.

The tensor-to-scalar ratio $r$ versus the spectral index $n_{\mathcal{R}}$ for the parameters region I of Fig.10 are show in Fig.11.
\begin{figure}
 \begin{minipage}[t]{0.49\linewidth} 
	\centering
	\includegraphics[width=.99\textwidth]{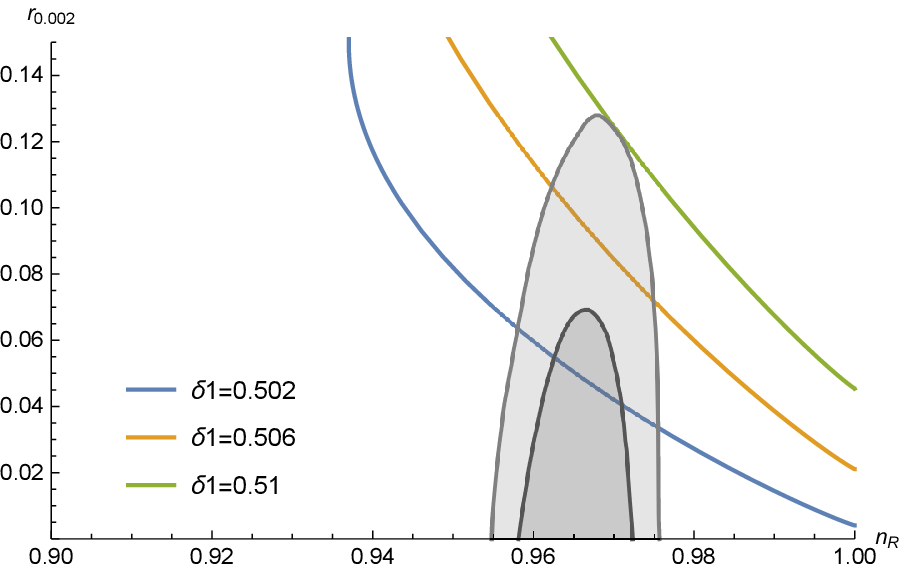}
	\label{fig:a} %
 \end{minipage}
 \begin{minipage}[t]{0.49\linewidth} 
	\centering
	\includegraphics[width=.99\textwidth]{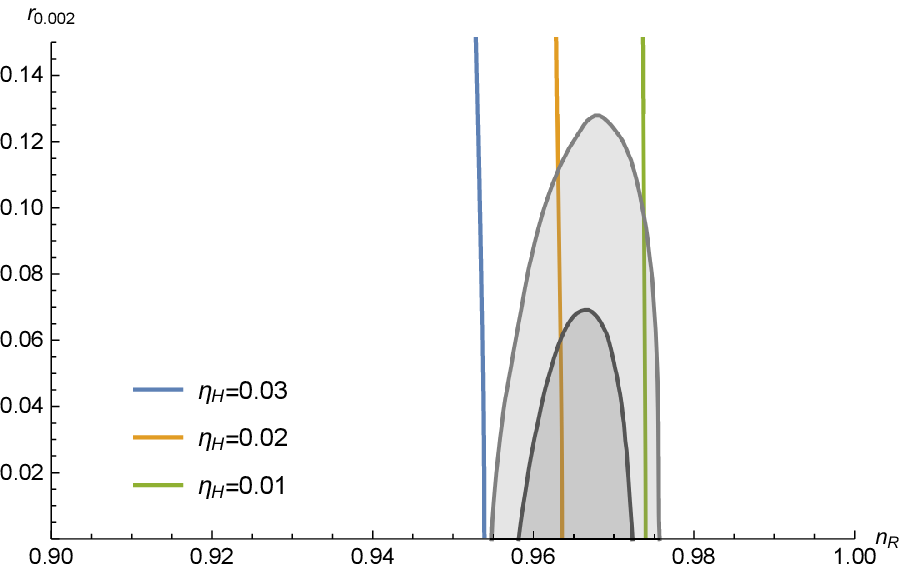}
	\label{fig:b}
 \end{minipage}
 \caption{ The $n_{\mathcal{R}}-r$ region predicted by the model for the parameter region I of Fig.10. }
\label{fig:1}
\end{figure}
 $\delta_1$ is taken as $\delta_1=0.502,0.506,0.51$ from left to right in the left panel, and the $n_{\mathcal{R}}-r$ dots go along the curves from right to left as $\eta_H$ increase.  The parameters $\eta_H$  taken as $\eta_H=0.03,0.02,0.01$ are show in the right panel, and this time the $n_{\mathcal{R}}-r$ dots go from bottom to top as $\delta_1$ increase.

In Fig.12, we show $n_{\mathcal{R}}$ versus $r$  for the parameters in region II of Fig.10.
\begin{figure}
 \begin{minipage}[t]{0.49\linewidth} 
	\centering
	\includegraphics[width=.99\textwidth]{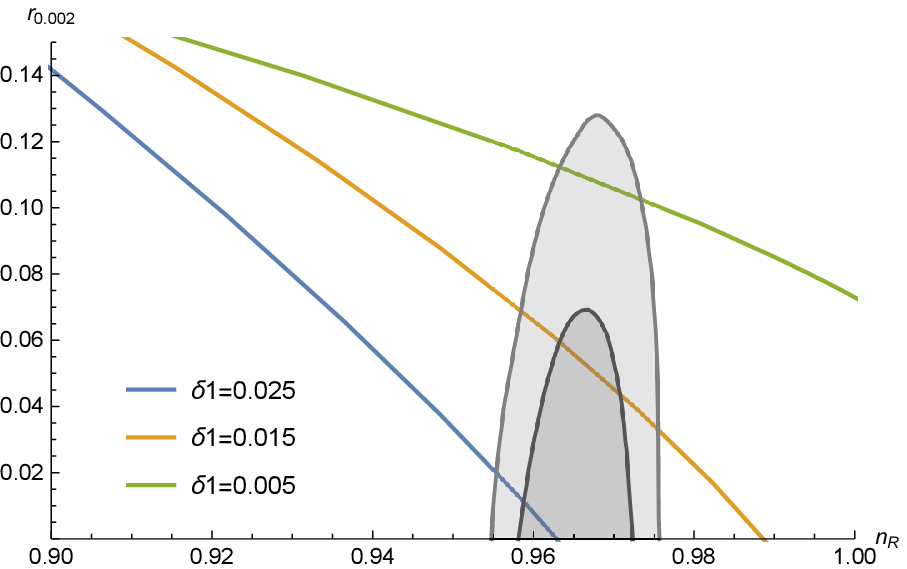}
	\label{fig:a} %
 \end{minipage}
 \begin{minipage}[t]{0.49\linewidth} 
	\centering
	\includegraphics[width=.99\textwidth]{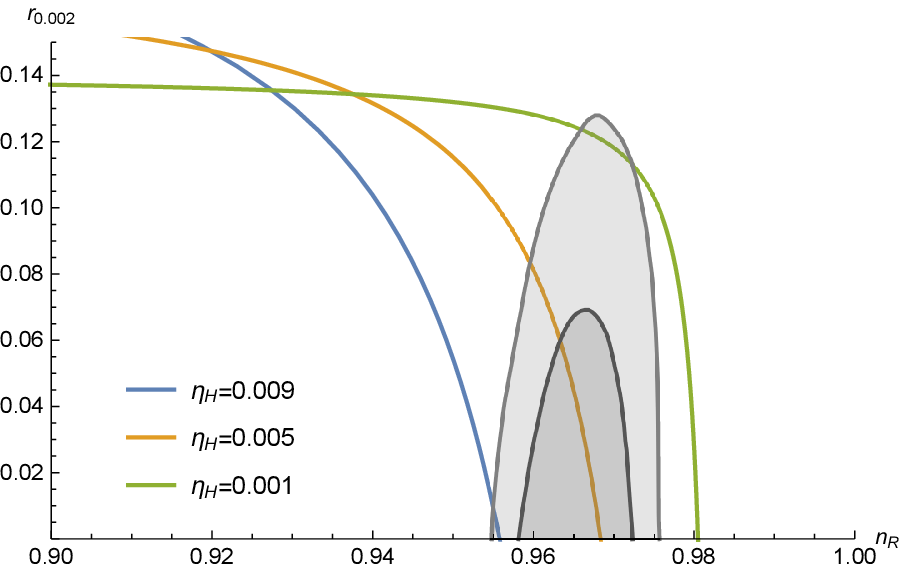}
	\label{fig:b}
 \end{minipage}
 \caption{ The $n_{\mathcal{R}}-r$ region predicted by the model for the parameter region II of Fig.10. }
\label{fig:1}
\end{figure}
Left panel: The parameter $\delta_1$ is taken as $\delta_1=0.025,0.015,0.005$ from left to right, and as $\eta_H$ increase, the $n_{\mathcal{R}}-r$ dots go along the curves from right to left. Right panel: The parameters $\eta_H$ is taken as $\eta_H=0.009,0.005,0.001$ from left to right, and as $\delta_1$ increase, the $n_{\mathcal{R}}-r$ dots go along the curves from left to right.

Similarly, the $n_{\mathcal{R}}$ versus $r$  for the parameters in region III of Fig.10 are show in Fig.13
\begin{figure}
 \begin{minipage}[t]{0.49\linewidth} 
	\centering
	\includegraphics[width=.99\textwidth]{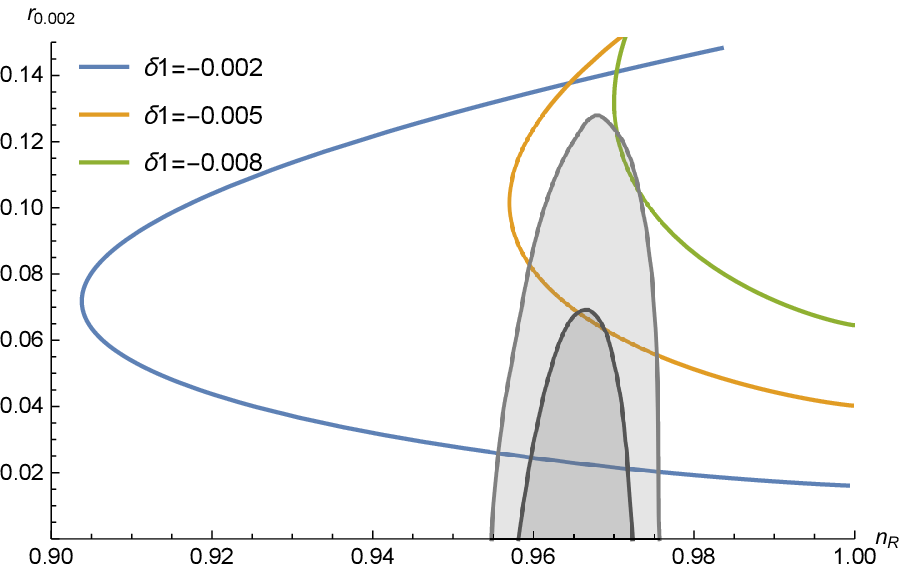}
	\label{fig:a} %
 \end{minipage}
 \begin{minipage}[t]{0.49\linewidth} 
	\centering
	\includegraphics[width=.99\textwidth]{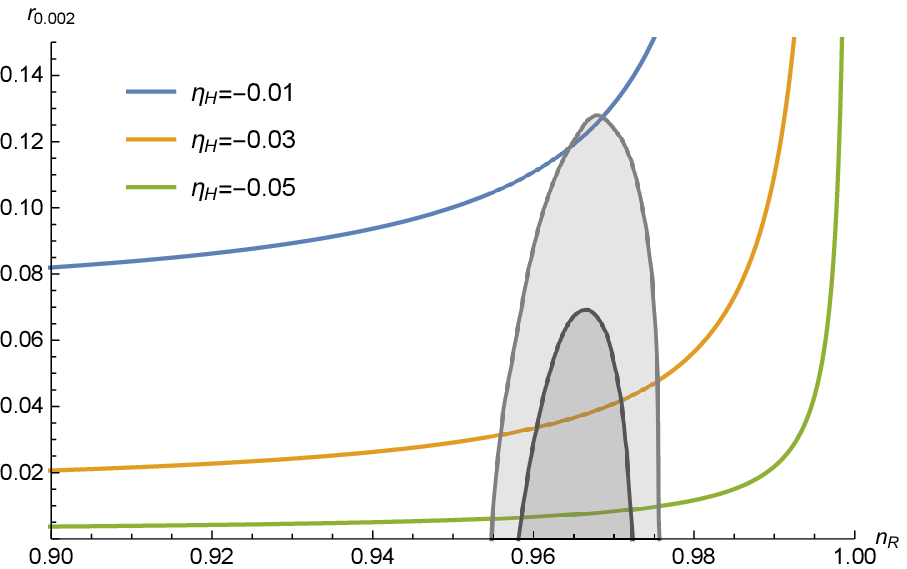}
	\label{fig:b}
 \end{minipage}
 \caption{ The $n_{\mathcal{R}}-r$ region predicted by the model for the parameter region III of Fig.10. }
\label{fig:1}
\end{figure}
In the left panel  The parameter is taken as $\delta_1=-0.002,-0.005,-0.008$ from left to right, and  the dots go along the curves to the top as $\eta_H$ increase.  The parameter $\eta_H$ is taken as $\eta_H=-0.01,-0.03,-0.05$ from left to right in the  right panel. In this time the $n_{\mathcal{R}}-r$ dots go from left to right as $\delta_1$ increase.

Finally, the $n_{\mathcal{R}}-r$ constraint for the parameters in region IV of Fig.10 are show in Fig.14.
\begin{figure}
 \begin{minipage}[t]{0.49\linewidth} 
	\centering
	\includegraphics[width=.99\textwidth]{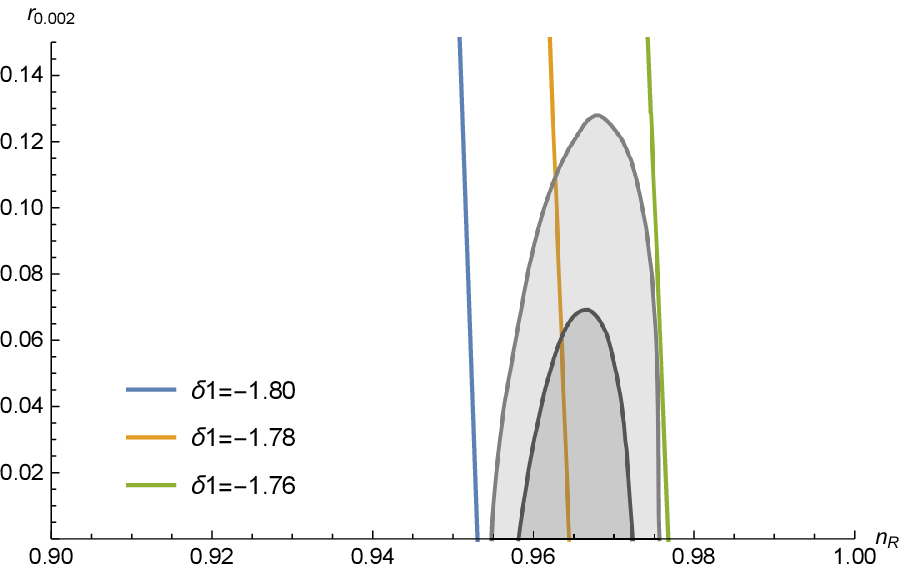}
	\label{fig:a} %
 \end{minipage}
 \begin{minipage}[t]{0.49\linewidth} 
	\centering
	\includegraphics[width=.99\textwidth]{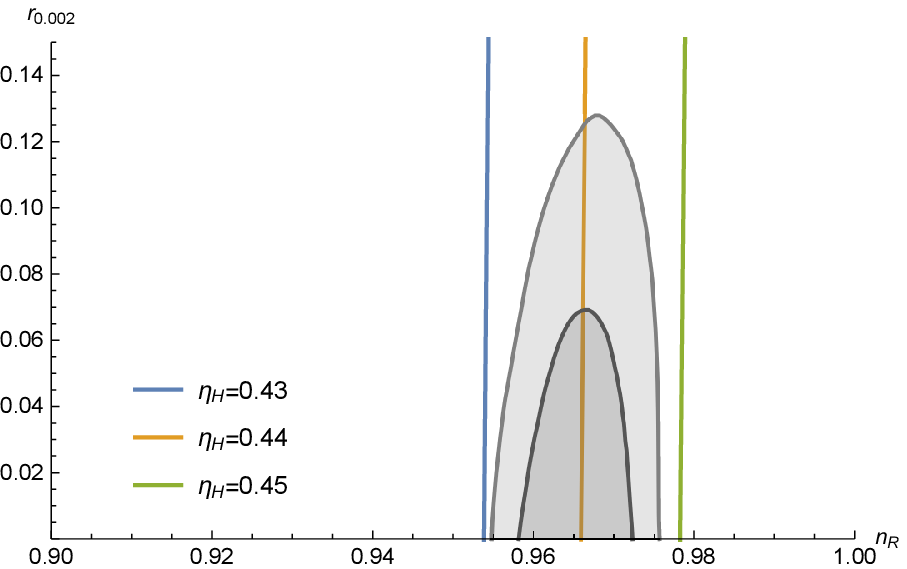}
	\label{fig:b}
 \end{minipage}
 \caption{ The $n_{\mathcal{R}}-r$ region predicted by the model for the parameter region IV of Fig.10. }
\label{fig:1}
\end{figure}
 with the parameters taken as $\delta_1=-1.8,-1.78,-1.76$ from left to right in the left, where the $n_{\mathcal{R}}-r$ dots go from bottom to top as $\eta_H$ increase.  The parameter in the right panels $\eta_H=0.43,0.44,0.45$ from left to right. In this time as $\delta_1$ increase, the dots go along the curves from bottom to top.

\section{Reconstruct the Potential \label{sec4}}
In this section, we shall find the corresponding scalar potential $V(\phi)$ and the GB coupling $\xi(\phi)$ for the constant-roll inflation. If the GB flow parameter $\delta_1$ is a constant, from the definition of flow parameters, we have $\delta_2=0$, then the background equations (2)and (3) can be written as
\begin{equation}
\omega \dot{\phi}^2=(2\epsilon_1-\delta_1-\delta_1 \epsilon_1){H^2},
\end{equation}
\begin{equation}
6 (1-\delta_1) H^2 =\omega \dot{\phi}^2+2V.
\end{equation}
In the following, we will discuss the models with $\epsilon_1=$constant, $\epsilon_2=$constant and $\eta_H=$constant, respectively.

\subsection{$\epsilon_1=$ constant}
In the model with $\epsilon_1=$ constant, the case with $\delta_1=0$ is ruled out by the observations, so we only interest in the case with $\delta_1=$constant but not zero.
Substitute (51) into the definition of the flow parameter $\epsilon_1$
\begin{equation}
\epsilon_{1}=-\frac{\dot{H}}{H^{2}}=-\frac{\dot{\phi}H_{,\phi}}{H^{2}},
\end{equation}
we obtain a first-order differential equation of the Hubble parameter and get two solutions
\begin{equation}
H(\phi)= c_1 e^{\pm \frac{i\epsilon_1\sqrt{\omega }\; \phi}{\sqrt{\delta_1-2 \epsilon_1+\delta_1 \epsilon_1 }}},
\end{equation}
with $c_1$ is a integration constant.
Substitute the Eqs.(54) and (51) into the Hamilton-Jacobi equation(52), we get the potential of the scalar field $\phi$
\begin{equation}
V(\phi)=V_0(6-5\delta_1-2\epsilon_1 +\delta_1 \epsilon_1)e^{\pm\frac{2 i\epsilon_1\sqrt{\omega }\; \phi}{\sqrt{\delta_1-2 \epsilon_1+\delta_1 \epsilon_1 }}}
\end{equation}
where the overall factor $V_0=c_1^2/2$ can be restricted by the amplitude of the primordial curvature perturbations.

For the GB coupling $\xi(\phi)$, we substitute the result of Hubble parameter (54) into the definition of GB flow parameter $\delta_{1}=4 \dot{\xi} H $ and get a differential equation of $\xi(\phi)$, the solutions are
\begin{equation}
\xi(\phi)=\frac{\delta_1}{16\epsilon_1 V_0}e^{\mp\frac{2 i\epsilon_1\sqrt{\omega }\; \phi}{\sqrt{\delta_1-2 \epsilon_1+\delta_1 \epsilon_1 }}}+\xi_0
\end{equation}
where $\xi_0$ is an integration constant represents the GB term with out coupling to $\phi$, which is no contribution to the result, and is always set to zero.
In addition, the coefficient $\omega$ in the expressions can be chosen the values $\pm1$ to ensure the results $H(\phi)$,$V(\phi)$,$\xi(\phi)$  are real functions and  generate reasonable inflation. For instance, if $\delta_1-2 \epsilon_1+\delta_1 \epsilon_1>0$,   $\omega$ should be chosen $-1$.

\subsection{$\epsilon_2=$ constant}
If the flow parameter $\epsilon_2$ is a constant, as discussed in section 3, the case with $\delta_1=0$ is also ruled out by the observations. For the case with $\delta_1\neq0$, the differential equation to the Hubble parameter is complicated and  no analytic solutions. However, for some special cases, we can get the approximate expressions. In the following, we focus on two special cases: $\epsilon_1\ll\delta_1$ and $\delta_1\ll\epsilon_1<1$.
\subsubsection{$\epsilon_1\ll\delta_1$}
 We first discuss the relation $\epsilon_1\ll\delta_1$, which can be satisfied near the beginning of inflation. In this case, the background equation (51) is approximated as
\begin{equation}
\omega \dot{\phi}^2=-\delta_1{H^2}.
\end{equation}
Substitute it into the definition of the flow parameter $\epsilon_2$
\begin{equation}
\quad \epsilon_{2}\equiv\frac{\dot{\epsilon}_{1}}{H \epsilon_{1}}=\frac{\dot{\phi}^{2} H_{, \phi \phi}+\ddot{\phi} H_{, \phi}}{\dot{\phi} H H_{\phi}}-\frac{2 \dot{\phi} H_{, \phi}}{H^{2}},
\end{equation}
and solve the second-order differential equation of $H$, we get the Hubble parameter  as a function of $\phi$
\begin{equation}
H(\phi)=c_2 e^{\pm\frac{i\;c_1 \sqrt{\delta_1} \text{exp}[\mp{\frac{i\;\epsilon_2 \sqrt{\omega }\; \phi }{\sqrt{\delta_1}}}]}{\epsilon_2 \sqrt{\omega } }},
\end{equation}
with $c_1$ and $c_2$ are integration constants.
Then combine the solution (59) with the background equations (51) and (52), we get the potential of the scalar field $\phi$
\begin{equation}
V(\phi)=V_0 (6-5 \delta_1) e^{\pm\frac{2i\;c_1 \sqrt{\delta_1} \text{exp}[\mp{\frac{i\;\epsilon_2 \sqrt{\omega }\; \phi }{\sqrt{\delta_1}}}]}{\epsilon_2 \sqrt{\omega } }},
\end{equation}
where  $V_0=c_2^2/2$ is the overall factor.

Combine of the definition $\delta_{1}=4 \dot{\xi} H $ and  the Hubble parameter (59), we get a differential equation the GB coupling $\xi(\phi)$, and the solutions are
\begin{equation}
\xi(\phi)=\frac{\delta 1 }{8 \epsilon_2 V_0}\text{Ei}\left(\mp\frac{2i\;c_1 \sqrt{\delta_1} \text{exp}[\mp{\frac{i\;\epsilon_2 \sqrt{\omega }\; \phi }{\sqrt{\delta_1}}}]}{\epsilon_2 \sqrt{\omega } }\right)+\xi_0,
\end{equation}
where $Ei(x)\equiv-\int_{-x}^{\infty } \frac{e^{-t}}{t} \, dt$ is the exponential integral function, and $\xi_0$ is a integration constant represent the GB term with out coupling to $\phi$.

\subsubsection{$\delta_1\ll\epsilon_1$}
The relation $\delta_1\ll\epsilon_1<1$ is likely to happen near the end of inflation, then the background equation (51) is approximated as
\begin{equation}
\omega \dot{\phi}^2=2\epsilon_1{H^2},
\end{equation}
combine with the definition of $\epsilon_1$, we get the relation
\begin{equation}
\omega\dot{\phi}=-2 H_{,\phi},
\end{equation}
which is the same as the standard slow-roll inflation without GB coupling.
Substitute (63) into the definition of the flow parameter $\epsilon_2$, Eq.(58),
we obtain the differential equation of the Hubble parameter and the  solution is
\begin{equation}
H(\phi)=c_{2} e^{c_{1} \phi-\frac{1}{8} \epsilon_{2} \omega \phi^{2}},
\end{equation}
with $c_1$ and $c_2$ are  integration constants.
The potential of the scalar field can be obtained by substitute the Eqs.(63) and (64) into (52), which can be written as
\begin{equation}
V(\phi)=V_0 \left(6-6 \delta_1-\frac{4 c_1^2}{\omega}+2c_1\epsilon_2\phi-\frac{1}{4}\epsilon_2^2\;\omega\phi^2\right)e^{2 c_{1} \phi-\frac{1}{4} \epsilon_{2}\omega\phi^{2}},
\end{equation}
with the overall factor $V_0=c_{2}^{2}/2$ .

Substitute the result of Hubble parameter (64) into the definition of GB flow parameter $\delta_{1}=4 \dot{\xi} H $, then we get the solution of GB coupling
\begin{equation}
\xi(\phi)=\frac{\delta_{1}}{8  \epsilon_{2} V_0} e^{-\frac{4 c_1^{2}}{\epsilon_{2}\omega}} \mathrm{Ei}\left[\frac{\left(\epsilon_{2}\omega \phi-4 c_{1}\right)^{2}}{4 \epsilon_{2}\omega}\right]+\xi_0.
\end{equation}

\subsection{$\eta_H=$ constant}

In this section, we reconstruct the  potential for the constant-roll inflation with $\eta_H$ and $\delta_1$ are constants. If $\delta_1=0$, the model recover to the case without GB coupling, which is discussed in other reference\cite{ref3}. Form the background equation, we get the relation
\begin{equation}
\omega\dot{\phi}=-2 H_{,\phi}
\end{equation}
Using the same method as in the previous subsections and combine the definition of $\eta_H$
\begin{equation}
\eta_H \equiv-\frac{\ddot{H}}{2H\dot{H}}=-\frac{\dot{\phi}^{2} H_{, \phi \phi}+\ddot{\phi} H_{, \phi}}{2\dot{\phi} H H_{\phi}},
\end{equation}
we get the Hubble parameter and  the scalar potential as functions of $\phi$
\begin{equation}
H(\phi)= c_1 e^{\frac{\sqrt{\eta_H} \sqrt{\omega} \phi}{\sqrt{2}}}+c_2 e^{-\frac{\sqrt{\eta_H} \sqrt{\omega} \phi}{\sqrt{2}}},
\end{equation}
\begin{equation}
V(\phi)= 2 c_1 c_2 (3+\eta_H)+(3-\eta_H)(c_1^2  e^{\sqrt{2\eta_H} \sqrt{ \omega } \phi}+c_2^2  e^{-\sqrt{2\eta_H} \sqrt{ \omega } \phi}),
\end{equation}
with $c_1$ and $c_2$ are  integration constants. The results are consistent with other reference\cite{ref3}.

Similarly, for the case with $\delta_1\neq0$, the differential equation to the Hubble parameter is no analytic solutions. And we discuss the approximate expressions in  two special cases: $\epsilon_1\ll\delta_1$ and $\delta_1\ll\epsilon_1<1$.

\subsubsection{$\epsilon_1\ll\delta_1$}
If $\epsilon_1\ll\delta_1$, combine the background equation (57) and (52) with the definition of parameter $\eta_H$ (68), we can get two solutions of the Hubble parameter $H(\phi)$ and the corresponding scalar potential $V(\phi)$
\begin{equation}
H(\phi)= c_{2} \left(e^{\pm\frac{2 i c_{1} \eta_H \sqrt{\omega}}{\sqrt{\delta_{1}}}}-e^{\pm\frac{2 i \eta_H \sqrt{\omega} \phi}{\sqrt{\delta_{1}}}}\right)^{\frac{1}{2}},
\end{equation}
\begin{equation}
V(\phi)=V_0\left(6-5 \delta_{1}\right)\left(e^{\pm\frac{2 i c_1 \eta_{H} \sqrt{\omega}}{\sqrt{\delta_{1}}}}-e^{\pm\frac{2 i \eta_{H} \sqrt{\omega} \phi}{\sqrt{\delta_{1}}}}\right),
\end{equation}
with $c_1$ and $c_2$ are  integration constants and $V_0=c_{2}^{2}/2$. Substitute the Hubble parameter (71) into the definition of GB flow parameter $\delta_{1}=4 \dot{\xi} H $, then the solution of GB coupling can be written as
\begin{equation}
\xi(\phi)=\xi_{0}+\frac{\delta_{1}}{16 V_0 \eta_{H}} e^{\frac{2 i c_{1} \eta_{H} \sqrt{\omega}}{\sqrt{\delta_{1}}}} \ln \left[e^{\frac{2 i c_{1} \eta_{H} \sqrt{\omega}}{\sqrt{\delta_{1}}}}-e^{\frac{2 i \eta_{H} \sqrt{\omega} \phi}{\sqrt{\delta_{1}}}}\right],
\end{equation}
for the $+$ case of (71), and for the $-$ case is
\begin{equation}
\xi(\phi)=\xi_{0}+\frac{\delta_{1}}{16 V_0 \eta_{H}} e^{-\frac{2 i c_{1} \eta_{H} \sqrt{\omega}}{\sqrt{\delta_{1}}}}\left(\ln \left[e^{\frac{2 i c_{1} \eta_{H} \sqrt{\omega}}{\sqrt{\delta_{1}}}}-e^{\frac{2 i \eta_{H} \sqrt{\omega} \phi}{\sqrt{\delta_{1}}}}\right]-\frac{2 i \eta_{H} \sqrt{\omega} \phi}{\sqrt{\delta_{1}}}\right).
\end{equation}

\subsubsection{$\delta_1\ll\epsilon_1$}
If $\delta_1\ll\epsilon_1<1$, using the same method, combine the relation (63) with the definition of parameter $\eta_H$ (68), we can get the solutions of the Hubble parameter $H(\phi)$
\begin{equation}
H(\phi)= c_1 e^{\frac{\sqrt{\eta_H} \sqrt{ \omega} \phi}{\sqrt{2}}}+c_2 e^{-\frac{\sqrt{\eta_H} \sqrt{ \omega} \phi}{\sqrt{2}}},
\end{equation}
then substitute it into the background equation (52), we get the scalar potential $V(\phi)$
\begin{equation}
V(\phi)= 2 c_1 c_2 (3+\eta_H-3 \delta_1)+(3-\eta_H-3 \delta_1)\left(c_1^2  e^{\sqrt{2\eta_H} \sqrt{ \omega } \phi}+c_2^2 e^{-\sqrt{2\eta_H} \sqrt{ \omega } \phi}\right),
\end{equation}
and the GB coupling $\xi(\phi)$ can be obtained by integer the definition of $\delta_1$
\begin{equation}
\xi(\phi)=\frac{\delta_{1} }{8 c_{1} c_{2} \eta_{H}}\tanh ^{-1}\left[\frac{c_{1} \mathrm{e}^{\sqrt{2 \eta_{H}} \sqrt{ \omega }\phi}}{c_{2}}\right]+\xi_{0}.
\end{equation}

\section{Summary \label{sec5}}
\indent\indent

In this paper, we discuss the constant-roll inflation in the model with the inflation field $\phi$ nonminimal couple to the GB term. Since the presence of a new degree of freedom from GB coupling $\xi(\phi)$, we consider a special case that the first GB flow parameter $\delta_1$ is a constant. Then we discuss the constant-roll inflation with constant $\epsilon_1$, constant $\epsilon_2$ and constant $\eta_H$, respectively. Using the Bessel function approximation, we present the mode equations of scalar and tensor perturbations and get the analytical expressions for the scalar and tensor power spectrum to the first order of $\epsilon_1$. We derive the scalar spectral index and the tensor to scalar ratio and constraint the parameter space by using the Planck 2018 results.
We first assume that $\epsilon_1$ is a constant, the case with $\delta_1=0$ is ruled out by the observations,  and in the case that $\delta_1=$ constant but not zero, we obtain two regions of parameter space and show the results on the $n_{\mathcal{R}}-r$ region.
Second, in the assumption with $\epsilon_2=$ constant, $\delta_1=0$ is the constant-roll inflation without GB coupling, which is also ruled out by the observations. If $\delta_1=$constant but not zero, we obtain three regions of parameter space.
Finally, we assume that $\eta_H=$ constant, the case with $\delta_1=0$ is the original constant-roll inflation discussed in \cite{ref3}. Here we update the constraint from Planck 2018 data, which is $-0.0163<\eta_H<-0.0006$.  In the case with $\delta_1=$ constant but not zero, we obtain four feasible regions of parameter space, and find that in some region, the chosen of GB flow parameter $\delta_1>1$ can also produce a nearly scale-invariant scalar power spectrum, which is different from the conventional slow-roll inflation with GB coupling.
In addition, we also find the analytical expressions for corresponding scalar potential of the constant-roll inflation and the nonminimum coupling $\xi(\phi)$ in some cases.

\begin{acknowledgments}
\indent\indent
We would like to thank Zong-Kuan Guo  for useful conversations. This work was supported by ``the National Natural Science Foundation of China'' (NNSFC) with Grant No. 11705133.
\end{acknowledgments}

\end{document}